\documentclass[letterpaper]{article} 
\usepackage{aaai2026}
\usepackage{times}  
\usepackage{helvet}  
\usepackage{courier}  
\usepackage[hyphens]{url}  
\usepackage{graphicx} 
\usepackage{natbib}  
\usepackage{caption} 
\usepackage{newfloat}
\usepackage{listings}
\DeclareCaptionStyle{ruled}{labelfont=normalfont,labelsep=colon,strut=off} 
\usepackage[linesnumbered,ruled,vlined]{algorithm2e}

\nocopyright

\usepackage{amsmath}
\usepackage{amssymb}
\usepackage{booktabs}
\usepackage{dsfont}
\usepackage{multirow}
\usepackage{placeins}
\usepackage{subcaption}

\usepackage{xspace}
\newcommand{\modelname}{ProtSAE\xspace}

\title{ProtSAE: Disentangling and Interpreting Protein Language Models via Semantically-Guided Sparse Autoencoders}
\author{
    Xiangyu Liu\textsuperscript{\rm 1},
    Haodi Lei\textsuperscript{\rm 1},
    Yi Liu\textsuperscript{\rm 1}, 
    Yang Liu\textsuperscript{\rm 1}, 
    Wei Hu\textsuperscript{\rm 1,2,}\thanks{Corresponding author}
}
\affiliations{
    \textsuperscript{\rm 1} State Key Laboratory for Novel Software Technology, Nanjing University, China\\
    \textsuperscript{\rm 2} National Institute of Healthcare Data Science, Nanjing University, China\\
    \{xyl.nju, haodilei, yiliu07.nju, yliu20.nju\}.nju@gmail.com, whu@nju.edu.cn
}

\usepackage{bibentry}

\begin{document}

\maketitle

\begin{abstract}
Sparse Autoencoder (SAE) has emerged as a powerful tool for mechanistic interpretability of large language models.
Recent works apply SAE to protein language models (PLMs), aiming to extract and analyze biologically meaningful features from their latent spaces. 
However, SAE suffers from semantic entanglement, where individual neurons often mix multiple nonlinear concepts, making it difficult to reliably interpret or manipulate model behaviors.
In this paper, we propose a semantically-guided SAE, called ProtSAE. 
Unlike existing SAE which requires annotation datasets to filter and interpret activations, we guide semantic disentanglement during training using both annotation datasets and domain knowledge to mitigate the effects of entangled attributes.
We design interpretability experiments showing that ProtSAE learns more biologically relevant and interpretable hidden features compared to previous methods.
Performance analyses further demonstrate that ProtSAE maintains high reconstruction fidelity while achieving better results in interpretable probing.
We also show the potential of ProtSAE in steering PLMs for downstream generation tasks.

\end{abstract}


\section{Introduction}
In recent years, protein language models (PLMs)~\cite{esm2, progen2} have developed rapidly and  been widely applied to downstream tasks including protein function prediction~\cite{protein_function_prediction}, structural modeling \cite{esmfold}, and protein design~\cite{protein_design}.
However, the internal mechanisms of PLMs remain largely unknown~\cite{plmsae}.

For protein engineering, it is important to understand how latent features map to biological concepts, such as binding pockets, post-translational modifications, or fold families.
Such analysis facilitates the identification of spurious correlations and biases, enhancing both performance and robustness of PLMs.
Moreover, it allows for the extraction of latent relationships among protein characteristics, providing meaningful insights that can inform and support biological research~\cite{esm2hidden_knowledge}.
Early works have attempted to analyze protein models, exploring the relationships between attention mechanisms and amino acids~\cite{bertology}, as well as identifying neurons associated with certain biological concepts~\cite{knowledge_neuron_in_plm}.


\begin{figure}[t]
    \centering
    \includegraphics[width=\columnwidth]{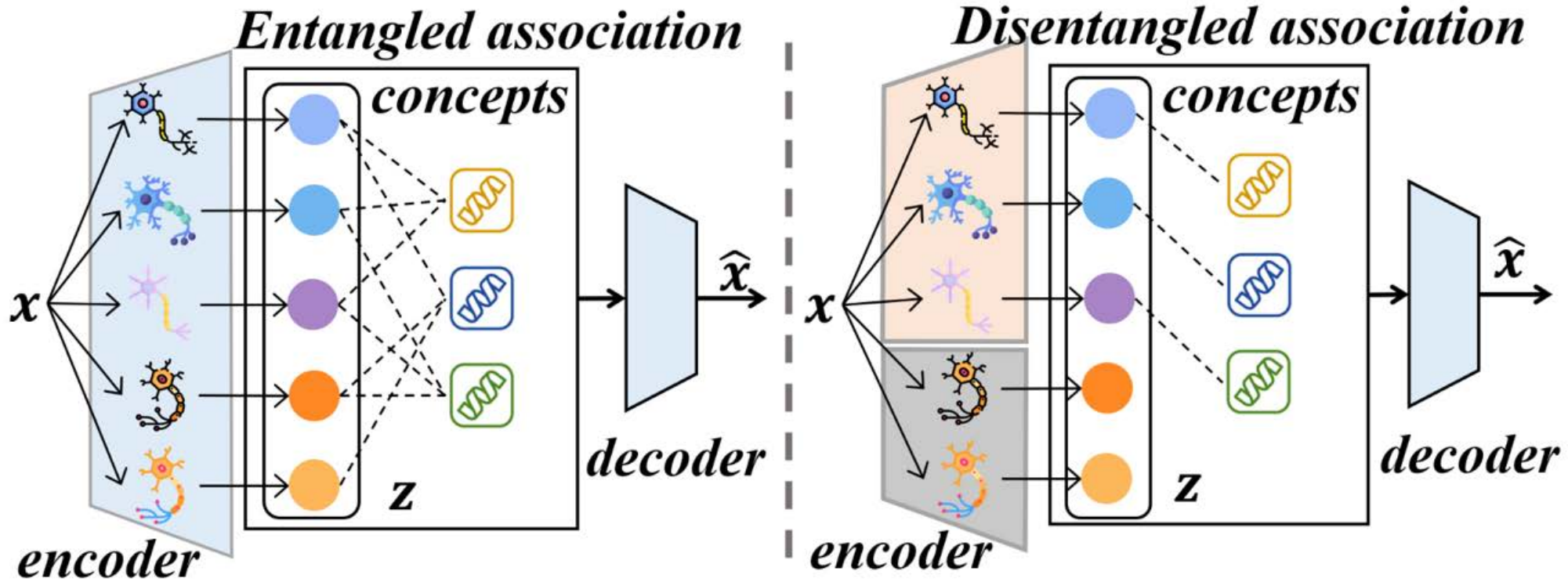}  
    \caption{Illustration of SAE semantic entanglement (left): individual neurons conflate multiple biological concepts, and semantic disentanglement (right): each defined neuron maps to a single biological concept.}
    \label{fig:intro_combined}
\end{figure}

Recent studies~\cite{interplm,plmsae,interprot} have begun applying sparse autoencoder (SAE) to PLMs and observed the emergence of features associated with various biological concepts.
SAE is an effective tool for understanding and explaining the internal representations of PLMs.
Based on the assumption of linear feature superposition~\cite{linear_superposition},
it decomposes the hidden representations to extract sparse features.
These features can be further analyzed for their correlations with specific concepts, enabling interpretability.
Furthermore, the sparse features can be selectively activated to steer the generation along the directions of relevant concepts.

A typical SAE requires an annotation dataset after training to interpret the learned features~\cite{interplm}. 
This annotation contains concepts of interest, and post-hoc correlation analysis is often performed to establish the relationship between the features in SAE and these concepts.
However, SAE suffers from the problem of semantic entanglement: \emph{individual neurons often conflate multiple concepts}~\cite{saeshifts}.
This entanglement results in ambiguous interpretations of the learned features.
As illustrated in Figure~\ref{fig:intro_combined}, each neuron is likely to be simultaneously associated with multiple, semantically divergent concepts.
Consequently, identifying the true meaning of the given feature and using it to steer the generation becomes challenging, undermining the interpretability of the model.

In this paper, we propose \modelname, which incorporates semantic guidance into the SAE training to disentangle semantic features.
First, we leverage the semantic annotations used for post-hoc interpretation of SAE features to constrain the relationship between defined activations and specific concepts during training.
We also use forced activations and feature rescaling to ensure that the defined activations effectively participate in reconstruction with high fidelity.
Second, considering the rich prior knowledge in the protein domain, where concepts are not mutually independent, we incorporate ELEmbeddings~\cite{elembeddings} to model potential logical constraints among concepts, e.g., subsumption and conjunction.
The constraints are integrated into the training process of \modelname to enhance the interpretability and semantic consistency of the learned features.

We construct interpretability experiments to demonstrate that \modelname effectively captures biologically meaningful features in PLMs, such as molecular functions, biological processes, and binding sites.
Compared with features annotated from the typical SAE, the defined neurons in \modelname learn more accurate, disentangled representations that are more tightly aligned with protein structures.
Furthermore, performance analyses show that \modelname preserves richer semantic information related to protein concepts.
Under varying levels of sparsity, it consistently achieves stronger performance on protein function prediction~\cite{deepgo2,deepgozero} while maintaining high reconstruction fidelity.
Finally, through targeted activation steering across various biological concepts, we demonstrate that the semantic features learned by \modelname can effectively guide PLM outputs toward desired functional outcomes—validating both the quality of the learned representations and their potential for precise model control.

The main contributions of this paper are listed as follows:
\begin{itemize}
\item We propose \modelname, a novel semantically-guided SAE that can disentangle complex protein features, yielding features in PLMs more strongly aligned with biological concepts.
See \url{https://github.com/nju-websoft/ProtSAE}.

\item We introduce protein domain knowledge into \modelname training to learn the logical constraints among concepts, and apply forced activations and feature rescaling to ensure that defined activations effectively participate in reconstruction with high fidelity.


\item We conduct extensive experiments and analyses.
Interpretability experiments demonstrate that \modelname captures more interpretable features that are closely aligned with biological concepts.
Detailed performance analyses show that \modelname consistently outperforms baselines across varying levels of sparsity while maintaining high reconstruction fidelity.
Steering experiments reveal that \modelname enables effective interventions across diverse biological concepts.
\end{itemize}

\section{Related Work}

\textbf{Mechanistic interpretability and SAE.}
Mechanistic interpretability aims to understand how neural networks produce outputs based on the internal algorithms that they have learned~\cite{MI}.
Previous works explore the computation subgraphs responsible for specific tasks~\cite{circuit1, circuit2, circuit3}, and analyze the behaviors within large language models (LLMs)~\cite{IOI,IOI2,IOI3}.
One prominent line of analysis focuses on identifying and studying sparse linear features within LLMs~\cite{function_vector}.
Based on the assumption of linear feature superposition~\cite{linear}, some works decompose language model activations and use them to intervene in the model's behavior~\cite{linear_superposition,codebook}.
Recent scaling efforts demonstrate the viability of SAE across LLMs, from Claude 3 Sonnet~\cite{claude} to GPT-4~\cite{topksae}, with extensions to multi-modal LLMs as well~\cite{vllmsae}.
Several works propose architectural improvements to SAE to mitigate feature shrinkage and improve reconstruction fidelity~\cite{shrinkage, gatedsae}.
Others focus on developing comprehensive evaluation frameworks for SAE~\cite{sae_classification} and exploring generating more informative explanations for activated features with additional datasets or LLMs~\cite{sae_explan1, mutual_information_sae}.

\smallskip
\noindent\textbf{Interpretability in PLMs.}
Early works in PLMs show that attention maps can capture structural and functional signals, including amino acid interactions~\cite{bertology}, protein contacts~\cite{protein_contact}, and functional sites like binding pockets and allosteric regions~\cite{allosteric, allosteric2}.
CB-pLM~\cite{concept_bottleneck} trains PLMs with a concept bottleneck layer for better understanding and controlling PLMs' generation.   
Recent studies explore how high-level conceptual knowledge is internally represented in the components of PLMs~\cite{protein_knowledge_neurons}.
SAE is used to decompose latent activations and reveal links between biological concepts and structural features~\cite{interplm,plmsae,interprot, gujral2025sparse}.
These studies also show that editing related activations can influence or steer protein sequence generation.
However, features from SAE often entangle multiple concepts, making interpretation unclear.
To address this, we introduce semantic guidance during the SAE training by linking specific activations to biological concepts, leading to more interpretable features.

\begin{figure*}[h]  
\centering  
\includegraphics[width=.95\textwidth]{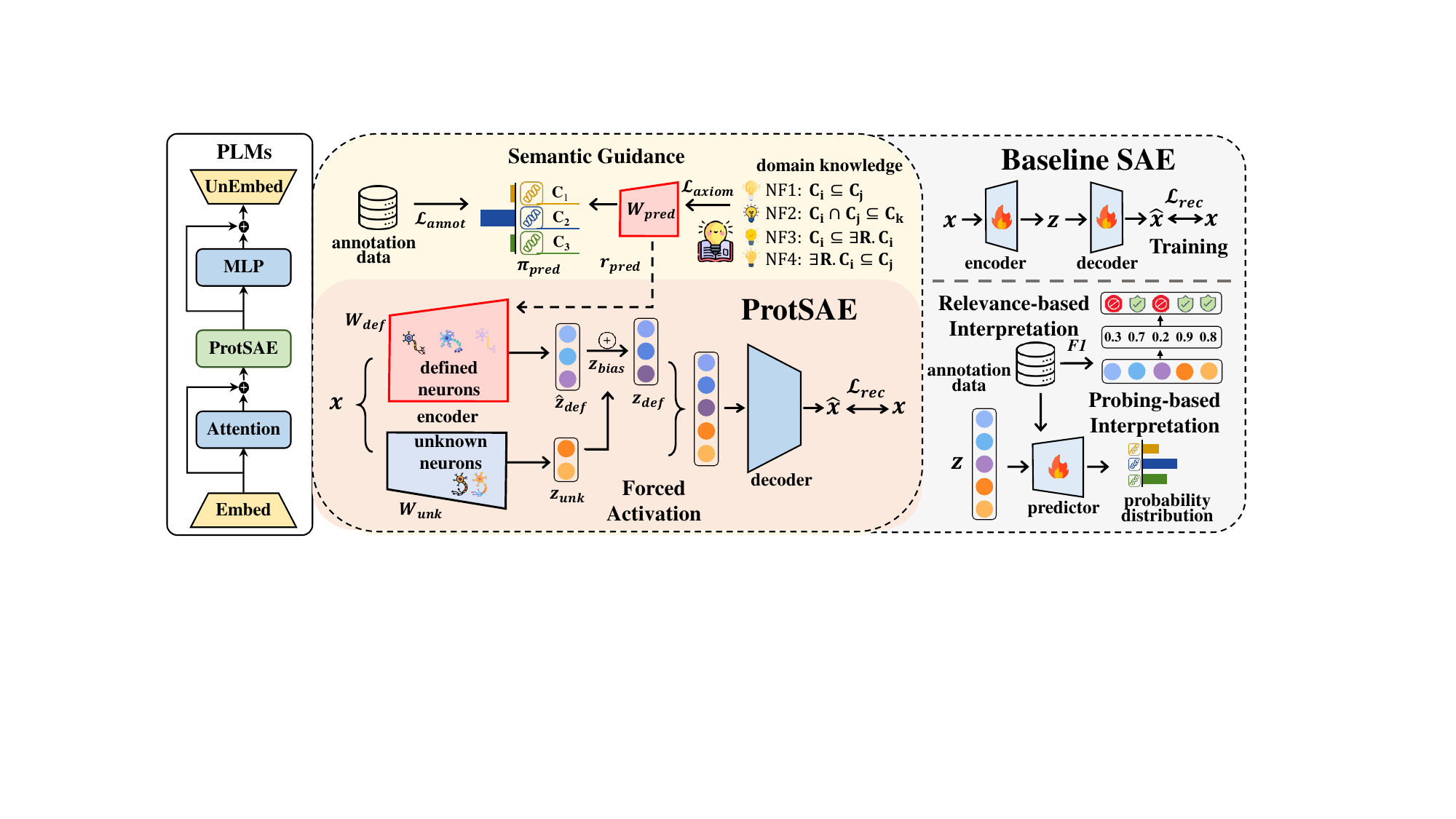}  
\caption{
An overview of \modelname (left) and the baseline SAE (right).
In the baseline SAE, annotation data is used post hoc to interpret learned features via relevance-based and probing-based methods.
In contrast, \modelname incorporates semantic guidance during training by leveraging annotation data and protein domain knowledge to achieve semantic disentanglement.
It also uses forced activations and feature rescaling to learn meaningful features while preserving reconstruction fidelity.
}  
\label{fig:framework}  
\end{figure*}

\section{Overview}
\textbf{SAE architectures.}
SAE is designed to learn sparse representations of high-dimensional inputs by encouraging only a small subset of neurons to be activated.
It is widely used to extract interpretable and localized features, particularly in the context of mechanistic interpretability for deep models.

Given an input vector $\mathbf{x} \in \mathbb{R}^d$, the encoder maps it to the latent activations $\mathbf{z} \in \mathbb{R}^n$ using a linear transformation followed by the ReLU activation:
\begin{align}
    \mathbf{z} &= \text{ReLU}(\mathbf{W}_{\text{enc}}(\mathbf{x} - \mathbf{b}_{\text{dec}}) + \mathbf{b}_{\text{enc}}),
\end{align}
where $\mathbf{W}_{\text{enc}} \in \mathbb{R}^{n \times d}$, $\mathbf{b}_{\text{enc}} \in \mathbb{R}^n$, and $\mathbf{b}_{\text{dec}}\in \mathbb{R}^d$ are learnable parameters. 
To enable sparsity in $\mathbf{z}$, $n$ is typically set much larger than $d$ ($n \gg d$).
The decoder reconstructs the input via a linear transformation:
\begin{align}
    \mathbf{\hat{x}} &= \mathbf{W}_{\text{dec}} \mathbf{z} + \mathbf{b}_{\text{dec}},
    \label{eq:reconstruction}
\end{align}
where $\mathbf{W}_{\text{dec}} \in \mathbb{R}^{d \times n}$.
The training objective encourages accurate reconstruction and sparsity in $\mathbf{z}$:
\begin{equation}
    \mathcal{L} = \|\mathbf{x}- \mathbf{\hat{x}}\|_2^2 + \lambda \|\mathbf{z}\|_1,
\end{equation}
where $ \|\mathbf{x} - \mathbf{\hat{x}}\|_2^2 $ is the reconstruction MSE loss, $\|\mathbf{z}\|_1$  imposes an L1 penalty to encourage sparsity, and \( \lambda \) is a tunable hyperparameter to balance the two terms.

\smallskip
\noindent\textbf{Interpreting SAE activations.}
As shown in Figure~\ref{fig:framework}, we follow prior works and use an annotation dataset that maps proteins to biological concepts to interpret the learned SAE features via two approaches:
(1) \textit{Relevance-based interpretation.}
Following previous work~\cite{plmsae}, we compute the activation levels of each feature across annotated proteins. 
Based on the annotation, we calculate a relevance score (e.g., F1-score) between a feature and a target concept, and use the relevant concept to interpret the feature.
Detailed formulations are described in the appendix.
(2) \textit{Probing-based interpretation.} We train linear probing classifiers on SAE activations using the annotation dataset~\cite{interplm,linear_probe}.
It aims to detect the presence of specific concepts within the learned features through supervised training.

\section{Method}

Conventionally, the encoder of SAE serves two purposes: 
(1) \textit{Determining which features should be active.}
To disentangle semantics in activated features, each defined neuron should be selectively activated only by proteins associated with a specific concept, while remaining inactive for unrelated protein sequences.
This constraint helps prevent entangled semantics within the same neuron.
Based on this intuition, we introduce semantic guidance from the annotation data and domain knowledge to learn such semantically selective activations during training.
(2) \textit{Estimating the magnitude of active features to support faithful reconstruction.} 
Although the magnitude of each active feature should be determined by the reconstruction training, we must ensure that the feature directions associated with the predefined concepts effectively contribute to reconstruction.
This guarantees that steering the encoder's activation yields consistent and interpretable effects on the model's behavior.

\subsection{Guiding SAE with Annotation Data}
We adopt the TopK-SAE as the backbone, where only the top-$K$ neurons with the highest activations are used in reconstruction.
We define the encoder as follows:
\begin{equation}
\mathbf{z} = \mathrm{TopK}\bigl(\mathbf{W}_{\rm enc} ( \mathbf{x} - \mathbf{b}_{\rm dec}) + \mathbf{b}_{\rm enc}\bigr),
\label{eq:encoder}
\end{equation}
where $\mathbf{W}_{\text{enc}} \in \mathbb{R}^{n \times d}$, $\mathbf{b}_{\text{enc}} \in \mathbb{R}^{n}$, and $\mathbf{b}_{\text{dec}} \in \mathbb{R}^{d}$.
\(\mathrm{TopK}(\cdot)\) retains only the top $K$ largest activations, zeroing out the rest.
Suppose that there are $m$ defined concepts of interest.
We aim for $m$ corresponding activations $\mathbf{z}_{\rm def}$ to accurately represent these concepts, while retaining the remaining $n-m$ activations $\mathbf{z}_{\rm unk}$ to capture unknown semantic concepts.
We partition the activation $\mathbf{z}$, weight matrix $\mathbf{W}_{\rm enc}$, and $\mathbf{b}_{\rm enc}$ into two components:
\begin{equation}
\mathbf{z} = \begin{bmatrix} \mathbf{z}_{\rm def} \\ \mathbf{z}_{\rm unk} \end{bmatrix}, 
\mathbf{W}_{\rm enc} = \begin{bmatrix} \mathbf{W}_{\rm def} \\ \mathbf{W}_{\rm unk} \end{bmatrix}, 
\mathbf{b}_{\rm enc} = \begin{bmatrix} \mathbf{b}_{\rm def} \\ \mathbf{b}_{\rm unk} \end{bmatrix},
\label{eq:partition}
\end{equation}
where $\mathbf{W}_{\rm def} \in \mathbb{R}^{m \times d}$ and $\mathbf{b}_{\rm def} \in \mathbb{R}^m$ corresponds to $m$ defined activations $\mathbf{z}_{\rm def}$ aligned with predefined concepts, 
$\mathbf{W}_{\rm unk} \in \mathbb{R}^{(n-m) \times d}$ and $\mathbf{b}_{\rm unk} \in \mathbb{R}^{n-m}$ capture the remaining activations $\mathbf{z}_{\rm unk}$.

\smallskip
\noindent\textbf{Semantic disentanglement.}
To guide the semantic disentanglement of specific activations, we introduce a concept predictor.
It learns to estimate the presence of each predefined concept from the input.
Let $\mathbf{W}_{\rm pred} \in \mathbb{R}^{m \times d}$ denote its weight matrix, the prediction probability of defined activations is computed as follows:
\begin{equation}
\pi_{\rm pred} = \sigma\bigl(\mathbf{W}_{\rm pred} (\mathbf{x} - \mathbf{b}_{\rm dec}) + \mathbf{b}_{\rm pred}\bigr) \in (0,1)^m,
\label{eq:prediction}
\end{equation}
where $\sigma(\cdot)$ is the sigmoid function and $\mathbf{b}_{\rm pred} \in \mathbb{R}^m$ is the prediction bias. 
We train this predictor on the available annotation data using a binary cross-entropy loss:
\begin{equation}
\mathcal{L}_{\rm annot} = \mathrm{CrossEntropy}(\pi_{\rm pred}, y),
\label{eq:annot}
\end{equation}
where $y \in \{0,1\}^m$ is the binary annotation vector indicating which semantic concepts are present.

We assume that  $\mathbf{W}_{\rm def}$ (used for reconstruction) and $\mathbf{W}_{\rm pred}$ (used for prediction) encode the same underlying semantic meanings. 
Thus, they should share the same projection directions. 
To achieve this, we treat $\mathbf{W}_{\rm def}$ as a rescaled version of $\mathbf{W}_{\rm pred}$, and tie their weights as follows:
\begin{equation}
\mathbf{W}_{\rm def} = \mathbf{W}_{\rm pred}^{\rm detach} \cdot \exp(\mathbf{r}_{\rm pred}),
\label{eq:weight_tying}
\end{equation}
where $\mathbf{r}_{\rm pred} \in \mathbb{R}^m$ is a learnable scaling vector and $\cdot$ denotes row-wise multiplication, where each row $i$ of $\mathbf{W}_{\rm pred}^{\rm detach}$ is scaled by $\exp(\mathbf{r}_{\rm pred}[i])$.
The \texttt{detach} indicates that gradients from the reconstruction loss are prevented from updating $\mathbf{W}_{\rm pred}$.
This formulation ensures that $\mathbf{W}_{\rm def}$ retains the semantic directionality learned from supervision, while its magnitude can adapt to improve reconstruction. 
The exponential guarantees positivity and allows smooth multiplicative modulation.

\smallskip
\noindent\textbf{Forced activation.}
Using the encoder, we compute the semantic and unsupervised activations as
\begin{align}
\mathbf{z}_{\rm unk} & = \mathrm{TopK}\bigl(\mathbf{W}_{\rm unk} (\mathbf{x} - \mathbf{b}_{\rm dec}) + \mathbf{b}_{\rm unk}\bigr), \\
\mathbf{\hat{z}}_{\rm def} & = \mathbf{W}_{\rm def} (\mathbf{x} - \mathbf{b}_{\rm dec}) + \mathbf{b}_{\rm def},
\label{eq:encoding}
\end{align}
where $\mathbf{\hat{z}}_{\rm def}$ denotes the pre-activation output for the defined concept neurons before any sparsity constraints.
In practice, we observe that the reconstruction tends to rely more on the entangled, unsupervised activations $\mathbf{z}_{\rm unk}$, which diminishes the contribution of concept-specific activations $\mathbf{z}_{\rm def}$. 
To mitigate this issue, we introduce a semantic bias that encourages the activations associated with predicted concepts to contribute more strongly to reconstruction:
\begin{align}
\mathbf{z}_{\rm bias} & = \mathds{1}_{\pi_{\rm pred}>0.5} \cdot \mathrm{ReLU}\bigl(\mathrm{mean}(\mathbf{z}_{\rm unk}) - \mathbf{\hat{z}}_{\rm def}\bigr), \\
\mathbf{z}_{\rm def} & = \mathbf{\hat{z}}_{\rm def} + \mathbf{z}_{\rm bias}.
\label{eq:bias}
\end{align}

Here, $\mathds{1}_{\pi_{\rm pred}>0.5} \in \{0,1\}^m$ denotes an indicator function, marking whether each semantic concept is predicted to be present (i.e., $\pi_{\rm pred} > 0.5$) and $\cdot$ denotes element-wise multiplication. 
For such concepts, we enforce the corresponding activation to be no less than the average activation of $\mathbf{z}_{\rm unk}$, preventing the model from ignoring semantically meaningful features during reconstruction. 
This bias effectively forces the features aligned with known semantics to participate in encoding, enhancing both interpretability and task relevance.

\subsection{Guiding SAE with Domain Knowledge}
\label{sec:sae-guidance}
In protein sequence modeling, biological concepts are often semantically interdependent. 
The relationship of some concepts can be defined with logical constraints including ``is-a'', ``part-of'', ``regulates'', and other relations.
These axioms establish a stable, expert-curated structure over biological knowledge, dictating how concepts relate and compose.
Therefore, aligning latent directions in SAE's hidden space with concept semantic relationships can enhance the detection and disentanglement of meaningful biological concepts.

To achieve this, we incorporate domain knowledge into SAE using ELEmbededings~\cite{elembeddings}.
It is an ontology representation learning method, which represents each concept as a hypersphere in the embedding space, and encodes logical axioms as constraints on the positions and relationships between these regions.
Given a concept $c_i$, the prediction probability of a protein $p$ can be modeled using ELEmbeddings as 
\begin{align}
\label{eq:elembedding-hyp}
y'_i = \sigma\left(f_\eta(p)^\top\cdot \big(f_\eta(hF) + f_\eta(c_i)\big) + r_\eta(c_i)\right),
\end{align}
where $f_\eta(\cdot)$ is the projection function into the semantic embedding space, $hF$ is the hasFunction relation, and $r_\eta(c_i) \in \mathbb{R}_{>0}$ is a learned radius bias.
In the appendix, we prove that Eq.~\eqref{eq:prediction} is structurally equivalent to Eq.~\eqref{eq:elembedding-hyp}.
Thus, from the perspective of ontology representation learning, the weight matrix $\mathbf{W}_{\mathrm{pred}}$ learned on the LLM latent space can be interpreted as an ontology embedding with relational biases.

We adopt four normalized axiom forms supported by ELEmbeddings, each corresponding to a specific type of logical relation commonly found in ontologies:
\begin{itemize}
\item \textbf{NF1.} Subclass axioms of the form $c_i \sqsubseteq c_j$, indicating that concept $c_i$ is a subclass of $c_j$.

\item  \textbf{NF2.} Conjunctive subclass axioms $c_i \sqcap c_j \sqsubseteq c_k$, stating that the intersection of concepts $c_i$ and $c_j$ is a subclass of $c_k$.

\item  \textbf{NF3.} Existential inclusion axioms of the form $c_i \sqsubseteq \exists R.c_j$, meaning that instances of $c_i$ are related via relation $R$ to some instance of $c_j$.

\item  \textbf{NF4.} Existential restriction axioms $\exists R.c_i \sqsubseteq c_j$, expressing that any entity related to an instance of $c_i$ via relation $R$ must belong to concept $c_j$.
\end{itemize}

These normalized forms serve as the basis for encoding ontological constraints as geometric relations within the embedding space.
The training loss is defined as
\begin{equation}
\mathcal{L}_{\rm axiom} = \mathcal{L}_{\rm NF1} + \mathcal{L}_{\rm NF2} + \mathcal{L}_{\rm NF3} + \mathcal{L}_{\rm NF4}.
\label{eq:axioms}
\end{equation}
where $L_1$ to $L_4$ represent the training losses of the four axioms under \modelname.
Appendix presents the detailed formulation of NF1 to NF4 and the derivation of the corresponding training loss.

\subsection{Training Strategy}  
The overall training objective combines the reconstruction loss, the supervised prediction loss, and a semantic regularization term guided by domain knowledge:
\begin{align}
\mathcal{L} 
=\; &\underbrace{\|\,\mathbf{\hat{x}} - \mathbf{x}\|_2^2}_{\mathcal{L}_{\rm rec}} 
+ \lambda_{\rm annot}\, \mathcal{L}_{\rm annot} 
+ \lambda_{\rm axiom}\, \mathcal{L}_{\rm axiom},
\end{align}
where $\mathbf{\hat{x}}$ is the reconstructed input defined in Eq.~\eqref{eq:reconstruction}. 
Here, $\mathcal{L}_{\rm rec}$ encourages faithful reconstruction of the input, 
$\mathcal{L}_{\rm annot}$ is the cross-entropy loss in Eq.~\eqref{eq:annot}, $\mathcal{L}_{\rm axiom}$ regularizes semantic alignment based on domain knowledge in Eq.~\eqref{eq:axioms}.
We use $\lambda_{\rm annot}$ and $\lambda_{\rm axiom}$ to weight these losses.
Appendix describes the detailed computation process.

\section{Experiments and Results}

\label{sect:exp}

\begin{figure*}[t]
    \centering
    \begin{subfigure}[t]{0.23\textwidth}
        \centering
        \includegraphics[width=\linewidth]{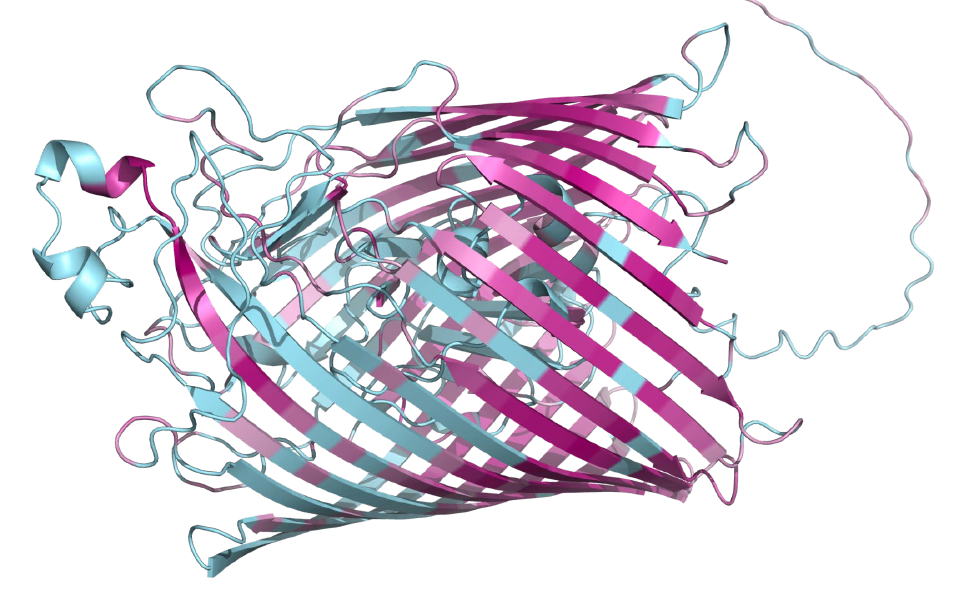}
        \caption{The feature related to \textit{iron ion binding} activated on the TonB-dependent receptor.}
        \label{fig:structure_2}
    \end{subfigure}
    \hspace{0.01\textwidth}
    \begin{subfigure}[t]{0.23\textwidth}
        \centering
        \includegraphics[width=\linewidth]{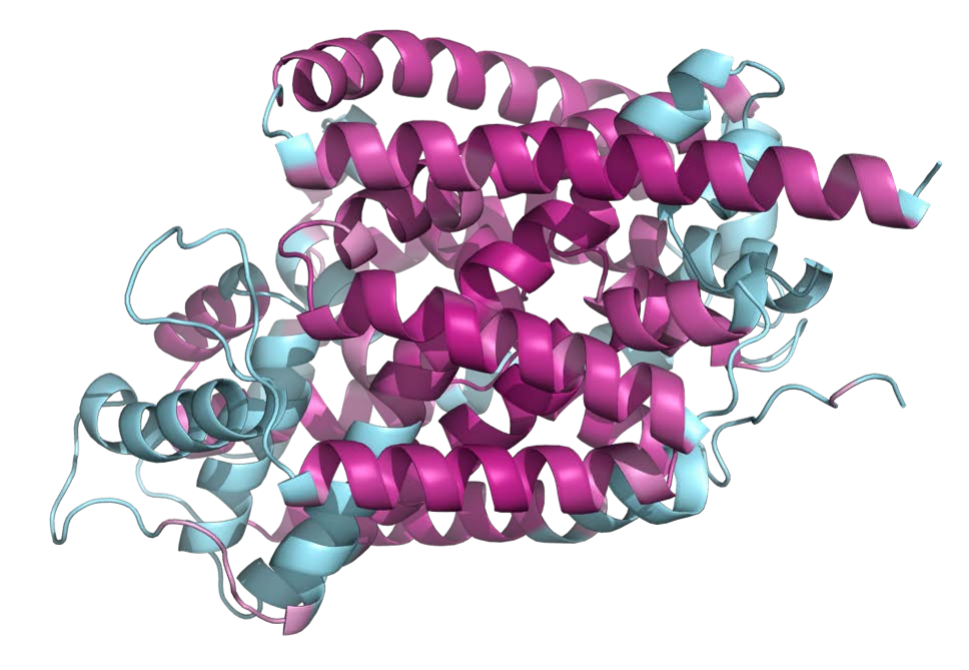}
        \caption{The feature related to \textit{sodium ion transport} activated on transmembrane segments.}
        \label{fig:structure_1}
    \end{subfigure}
    \hspace{0.01\textwidth}
    \begin{subfigure}[t]{0.23\textwidth}
        \centering
        \includegraphics[width=\linewidth]
        {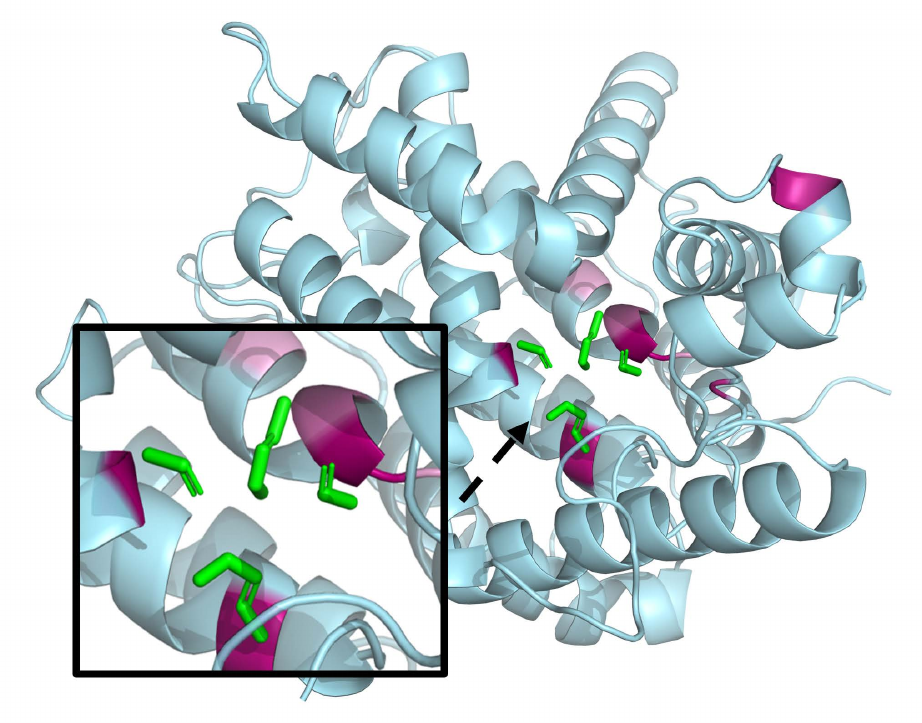}
        \caption{Interpreting $\text{Mn}^{2+}$ ion binding sites with an activated feature.}
        \label{fig:mn_binding_site}
    \end{subfigure}
    \hspace{0.01\textwidth}
    \begin{subfigure}[t]{0.23\textwidth}
        \centering
        \includegraphics[width=\linewidth]{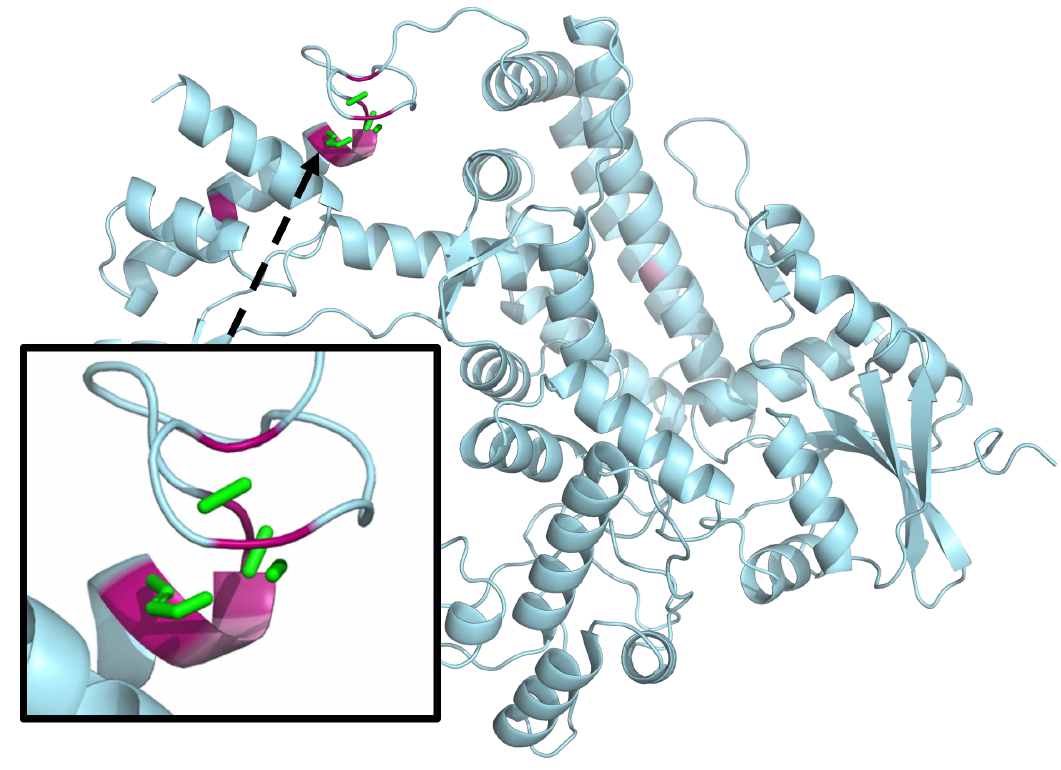}
        \caption{Interpreting $\text{Zn}^{2+}$ ion binding sites with an activated feature.}
        \label{fig:zn_binding_site}
    \end{subfigure}
    \caption{ Interpretability visualization shows that \modelname reveals semantic alignment between learned features and protein structures, including functional regions and ion binding sites.
    We use red intensity to indicate feature activation strength, and green sticks to mark ground truth binding sites.}
    \label{fig:combined_results}
\end{figure*}

\subsection{Experiment Setup}
\textbf{Dataset and baselines.}
We construct the annotation data from protein function prediction datasets~\cite{deepgo2}. 
The protein function prediction datasets contains 77,647 proteins extracted from UniProtKB/Swiss-Prot.
The annotated concepts can be categorized into three sub-ontologies: molecular function (MFO), biological process (BPO), and cellular component (CCO).
Furthermore, we use the ion binding sites dataset~\cite{metal_ion_binding} as another annotation dataset, which covers four biologically relevant ion types: Zn$^{2+}$, Ca$^{2+}$, Mg$^{2+}$, and Mn$^{2+}$.
We compare \modelname with widely adopted SAE baselines, including Naive SAE, Gated SAE~\cite{gatedsae}, and TopK SAE~\cite{topksae}, in terms of both interpretability experiments and performance analyses.
In the probing-based interpretation experiments, we further compare linear probing on the PLM hidden representations and the dictionary learning method SpLiCE~\cite{splice}.
Detailed datasets and baseline settings are included in the appendix.

\begin{figure}
    \centering
    \begin{subfigure}[t]{0.34\linewidth}
        \centering
        \includegraphics[width=\linewidth]{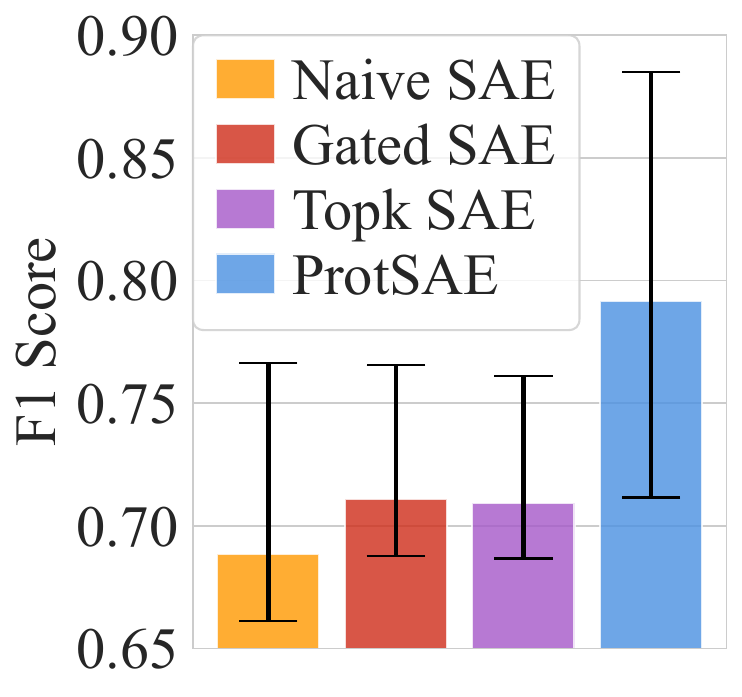}
        \caption{F1-score (mean, max, min) of top-10 activations. All results are reported as averages over 15 concepts.}
        \label{fig:mean_summary}
    \end{subfigure}
    \hfill
    \begin{subfigure}[t]{0.64\linewidth}
        \centering
        \includegraphics[width=\linewidth]{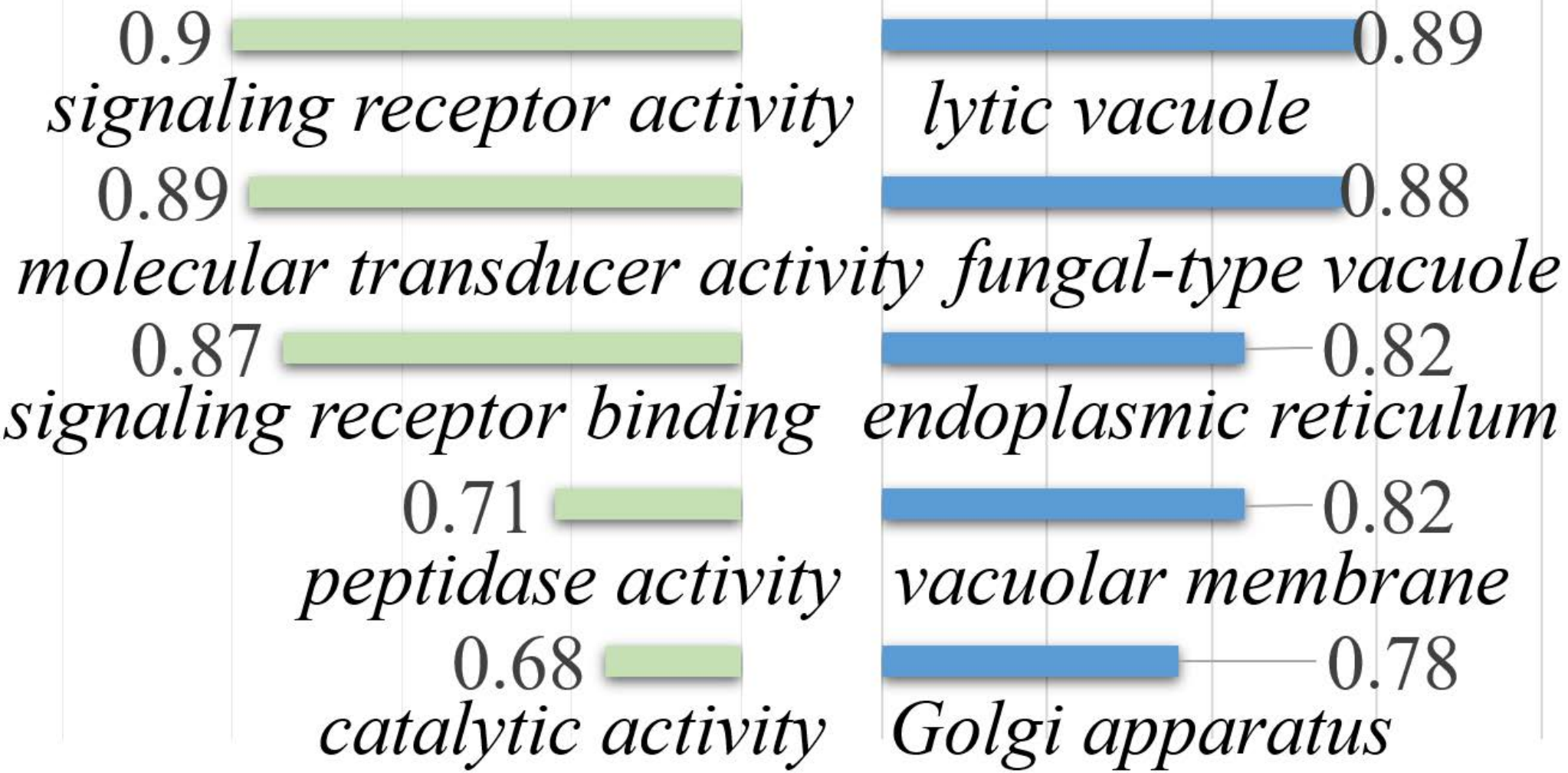}
        \caption{Top-5 activated concepts related to \textit{receptor ligand activity }~(left) and \textit{storage vacuole}~(right)}
        \label{fig:related_features}
    \end{subfigure}
    \caption{Comparison of relevance-based interpretation}
    \label{fig:relevance_based_interpretation}
\end{figure}

\smallskip
\noindent\textbf{Evaluation metrics.}
For relevance-based interpretation, we use F1-score to evaluate the relevance between neurons and biological concepts.
To evaluate the probing-based interpretation, we employ standard metrics from the protein function prediction benchmark~\cite{deepgo2} including AUPR, AUC, maximum protein-centric F-measure ($F_{\text{max}}$), and minimum semantic distance ($S_{\text{min}}$).
We use Loss Recovered~\cite{gatedsae} to assess the reconstruction fidelity on varying sparsity.
For intervention, we evaluate structural similarity using Template Modeling score (TM-score)~\cite{tm-score} and Root Mean Square Distance (RMSD)~\cite{rmsd}.
Details are described in the appendix.

\smallskip
\noindent\textbf{Implementation.}
We train all SAE on the internal activations of ESM2-15B, with $5\text{e}^{-4}$ learning rate, 12,800 batch size, and 25,000 steps. For \modelname and TopK SAE, the number of active neurons $K$ is varied in \{50, 100, 500, 1000\}. Gated SAE is tuned with L1 coefficients in \{$1.5\text{e}^{-4}$, $2\text{e}^{-4}$, $3\text{e}^{-4}$, $4\text{e}^{-4}$, $5\text{e}^{-4}$\}, and Naive SAE in \{$8\text{e}^{-5}$, $6\text{e}^{-5}$, $2\text{e}^{-4}$, $3\text{e}^{-4}$, $4\text{e}^{-4}$\}. 
All SAE activation width is set to 40,000 for BPO, 30,000 for MFO and CCO, and 10,000 for the ion binding-site dataset.
$\lambda_{\text{annot}}$ and $\lambda_{\text{axiom}}$ are fixed at 1.
Experiments are run on four NVIDIA A800 GPUs.

\subsection{Interpretability Experiments}

\textbf{Interpretability visualization.}
Figure~\ref{fig:combined_results} visualizes the features learned by \modelname that are aligned with biological concepts and demonstrates their utility in interpreting and exploring the semantics of specific protein structural elements.
Warmer colors (e.g., red) indicate stronger activation of the feature at that amino acid.
The positions of the true binding sites are marked by green sticks.
In Figure~\ref{fig:structure_2}, we examine activations associated with the concept \textit{iron ion binding} on protein P06971.
We observe strong activation in regions corresponding to the \textit{TonB-dependent receptor} structure, which is known to be tightly associated with the recognition and transport of $\mathrm{Fe}^{3+}$ ions.
In Figure~\ref{fig:structure_1}, the feature related to \textit{sodium ion transport} is highly activated on the transmembrane segments of protein O67854, suggesting that certain $\alpha$-helical transmembrane regions may play a crucial role in sodium ion transport.
Furthermore, Figures~\ref{fig:mn_binding_site} and \ref{fig:zn_binding_site} show that features related to \textit{metal ion binding sites} can highlight binding sites in proteins. 

\smallskip
\noindent\textbf{Relevance-based interpretation evaluation.}
In this experiment, we evaluate whether \modelname can effectively identify features related to specific concepts.
Following the previous work~\cite{interplm}, we extract relevant sequences from UniProtKB,\footnote{\url{https://www.uniprot.org/}} and construct a validation set of 5,000 sequences for each of 15 GO terms along with an equal number of unrelated sequences used as negative examples.
We compute the activation of each feature with respect to a given concept and report the top-10 features ranked by F1-score.
A higher F1-score indicates a stronger correlation between the feature and the concept.

\modelname identifies features that are more semantically aligned with the target concepts.
Figure~\ref{fig:mean_summary} shows the average results of 15 concepts.
Compared to the baselines, \modelname demonstrates significantly stronger relevance in both the mean and maximum activation levels.
This suggests that the incorporation of semantic guidance during training enables \modelname to effectively disentangle semantic signals, and learns more accurate concept-related features.
We further present additional highly activated features that show strong relevance to the validation set.
As shown in Figure~\ref{fig:related_features}, these features exhibit close functional or structural relationships with the target concept.
For example, in the case of \textit{storage vacuole}, the model activates structurally similar vacuole types such as \textit{lytic vacuole} and \textit{fungal-type vacuole}, as well as membrane-associated components like \textit{vacuolar membrane} and \textit{Golgi apparatus}.

\smallskip
\noindent\textbf{Probing-based interpretation evaluation.}
To better evaluate the capability of probing-based interpretation, we conduct protein function prediction across datasets from three ontologies.
The averaged results are summarized in Table~\ref{tab:main_result}.
For a fair comparison, all SAE-based methods are evaluated under the same sparsity level.
Notably, \modelname consistently outperforms all SAE baselines and the dictionary learning method  SpLiCE across all evaluation metrics, and achieves comparable performance to linear probing on the hidden representations of PLMs.
It suggests that \modelname, with semantic guidance, encourages the model to attend more effectively to the biological concepts during training. 
As a result, it mitigates the semantic loss that may occur during the SAE training, thereby achieving performance comparable to direct linear probing on PLMs.

\begin{table}
\centering
{\small
\begin{tabular}{lcccc}
\toprule
Method & $F_{\max}\uparrow$ & $S_{\min}\downarrow$ & AUPR$\uparrow$ & AUC$\uparrow$ \\
\midrule
SpLiCE      & .417 & 23.4 & .360 & .329 \\    
Naive SAE   & .421 & 23.3 & .340 & .511 \\
Gated SAE   & .441 & 22.7 & .368 & .533 \\
TopK SAE    & .444 & 22.7 & .379 & .565 \\
Linear Probe& \underline{.537} & \textbf{20.9} & \textbf{.522} & \underline{.751} \\  
\modelname  & \textbf{.579} & \textbf{20.9} & \underline{.487} & \textbf{.797}   \\
\bottomrule
\end{tabular}}
\caption{Average performance across three datasets on probing-based interpretation}
\label{tab:main_result}
\end{table}

\subsection{Performance Analyses}

\begin{figure}
    \centering
    \begin{subfigure}[t]{0.49\linewidth}
        \centering
        \includegraphics[width=\linewidth]{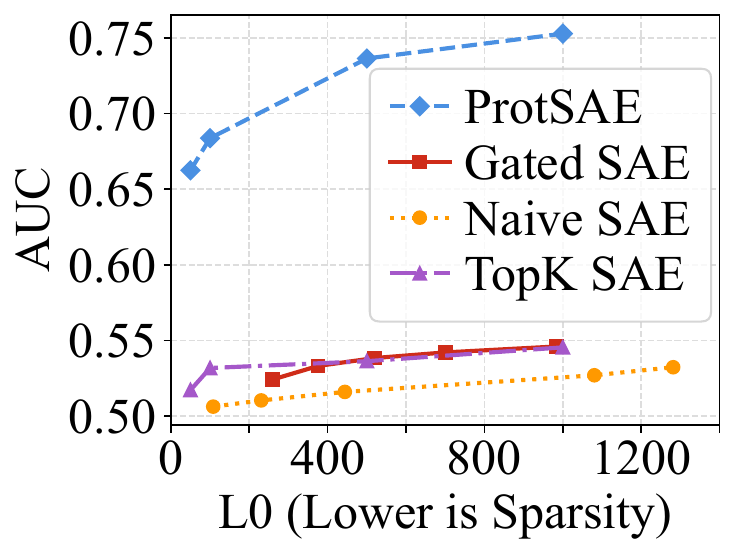}
        \caption{AUC under different sparsity}
        \label{fig:analysis_auc}
    \end{subfigure}
    \hfill
    \begin{subfigure}[t]{0.49\linewidth}
        \centering
        \includegraphics[width=\linewidth]{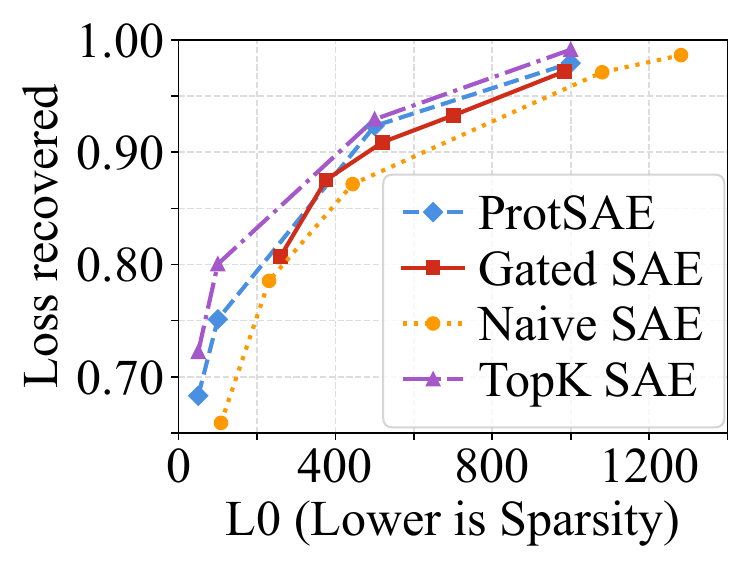}
        \caption{Loss Recovered under different sparsity}
        \label{fig:analysis_recovered}
    \end{subfigure}
    \caption{Performance comparison under different sparsity on the BPO dataset}
    \label{fig:analysis_result}
\end{figure}

\textbf{Performance across different sparsity.}
Figure~\ref{fig:analysis_result} illustrates the effect of sparsity on model performance.
We assess the reconstruction fidelity using the metrics of \emph{Loss Recovered}, which measures the proportion of the original PLM loss that can be recovered using SAE. 
The left subfigure shows the AUC performance under varying levels of sparsity.
\modelname consistently outperforms all baselines, indicating its ability to preserve semantics relevant to predefined concepts even under high sparsity, thereby achieving superior predictive performance.
The right subfigure shows the trend of reconstruction fidelity as sparsity increases.
Compared to other SAE variants, \modelname maintains comparable reconstruction quality, demonstrating its effectiveness in decomposing semantic concepts while faithfully preserving the original latent representations from PLMs.
Detailed results on all datasets are provided in the appendix.

\smallskip
\noindent\textbf{Ablation study.}
We conduct an ablation study of \modelname by removing key components and retraining the model under various sparsity levels.
Figure~\ref{fig:ablation_result} depicts the ablation results.
We discuss the effect of each component below:

\begin{figure*}[!ht]  
\centering  
\includegraphics[width=.95\textwidth]{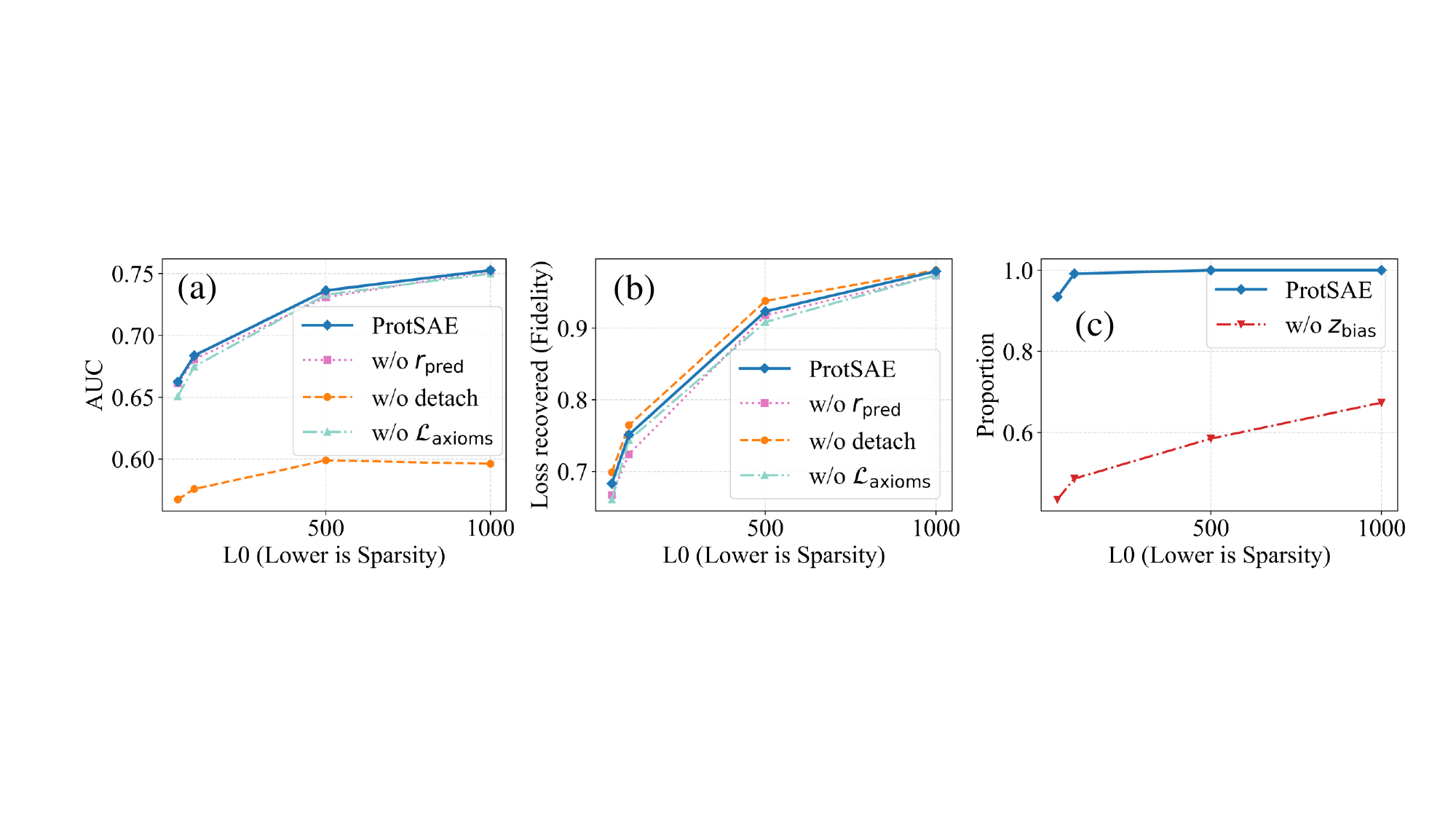}  
\caption{
Ablation results on (a) AUC, (b) Loss Recovered, and (c) reconstruction proportion of predicted activations w.r.t.  $L_0$
}  
\label{fig:ablation_result}  
\end{figure*}

\begin{itemize}
\item Without \texttt{detach}.
In this variant, we no longer detach $\mathbf{W}_{\text{pred}}$ when constructing $\mathbf{W}_{\text{def}}$, allowing the gradients from $\mathcal{L}_{\rm rec}$ to update $\mathbf{W}_{\text{pred}}$ directly.
So, the defined activations are now updated not only by the supervised data, but also by the reconstruction objective.
While this slightly improves reconstruction fidelity, it leads to a dramatic decrease in AUC.
This suggests that allowing $\mathcal{L}_{\rm rec}$ to influence $\mathbf{W}_{\text{pred}}$ introduces entangled or ambiguous semantics into the defined activations, thereby degrading precision and interpretability.

\item Removing $\mathcal{L}_{\text{axiom}}$.
When the axiom learning component based on ELEmbeddings is removed, the training of $\mathbf{W}_{\text{pred}}$ relies solely on the supervised data. 
This leads to a clear degradation in both AUC and reconstruction fidelity, highlighting the importance of modeling complex concept relationships through axioms to capture the intricate semantic structure of protein functions.

\item Without $\mathbf{z}_{\text{bias}}$.
In the right subfigure of Figure~\ref{fig:ablation_result}, we report the proportion of defined activations predicted as active that are indeed used during decoding.
With $\mathbf{z}_{\text{bias}}$, nearly all predicted activations participate in reconstruction, indicating that $\mathbf{z}_{\text{def}}$ holds strong potential for steering.
Removing $\mathbf{z}_{\text{bias}}$ reduces this proportion, weakening the alignment between prediction and actual activation.

\item Without $\mathbf{r}_{\text{pred}}$.
By setting the scaling parameter $\mathbf{r}_{\text{pred}}$ in Eq.~(\ref{eq:weight_tying}) to zero, $\mathbf{W}_{\text{def}}$ and $\mathbf{W}_{\text{pred}}$ become identical.
This modification causes drops in both AUC and reconstruction fidelity.
The removal of $\mathbf{r}_{\text{pred}}$ hinders the model's ability to learn feature magnitudes.
\end{itemize}





\subsection{Steering Experiment}


\begin{figure}[t]
    \centering
    \includegraphics[width=\columnwidth]{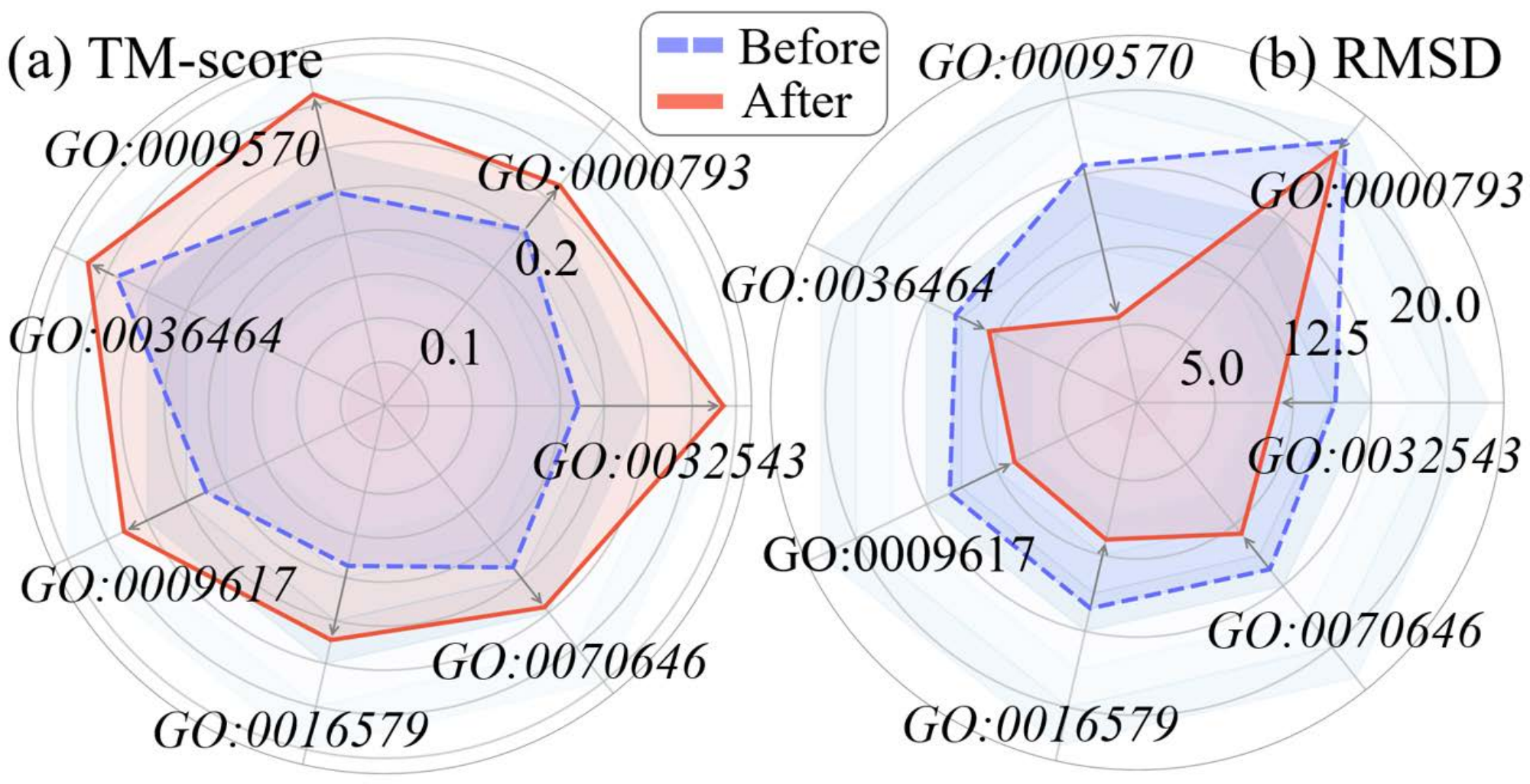}  
    \caption{Effect of TM-score (left) and RMSD (right) before and after intervention}
    \label{fig:intervention}
\end{figure}

\textbf{Concept intervention.}
We conduct a steering experiment across various biological concepts to evaluate whether \modelname can effectively steer PLM's generation based on the learned concept-specific features. 
For each of seven selected concept-related sequences, we mask 50\% of the tokens and compare the reconstructions generated before and after intervention. 
Following previous works~\cite{taxdiff,prollama,CtrlProt}, we use TM-score and RMSD to measure structural similarity between the generated sequences and the natural proteins with the target concept, in order to assess whether the intervention can guide the model's generation aligned with the desired concept.
We use pLDDT to evaluate the structure stability.

As shown in Figure~\ref{fig:intervention}, TM-scores significantly increase while RMSD decreases after intervention, indicating improved structural alignment with the target concepts. 
Furthermore, the appendix includes detailed results and highlights significant improvements in the pLDDT scores of the generated proteins.
These results suggest that \modelname successfully stores concept-aligned representations in its learned dictionary, and activating these features during generation enables the PLM to produce structurally stable proteins that better reflect the semantics of the desired concept.

\begin{figure}[t]
    \centering
    \includegraphics[width=\columnwidth]{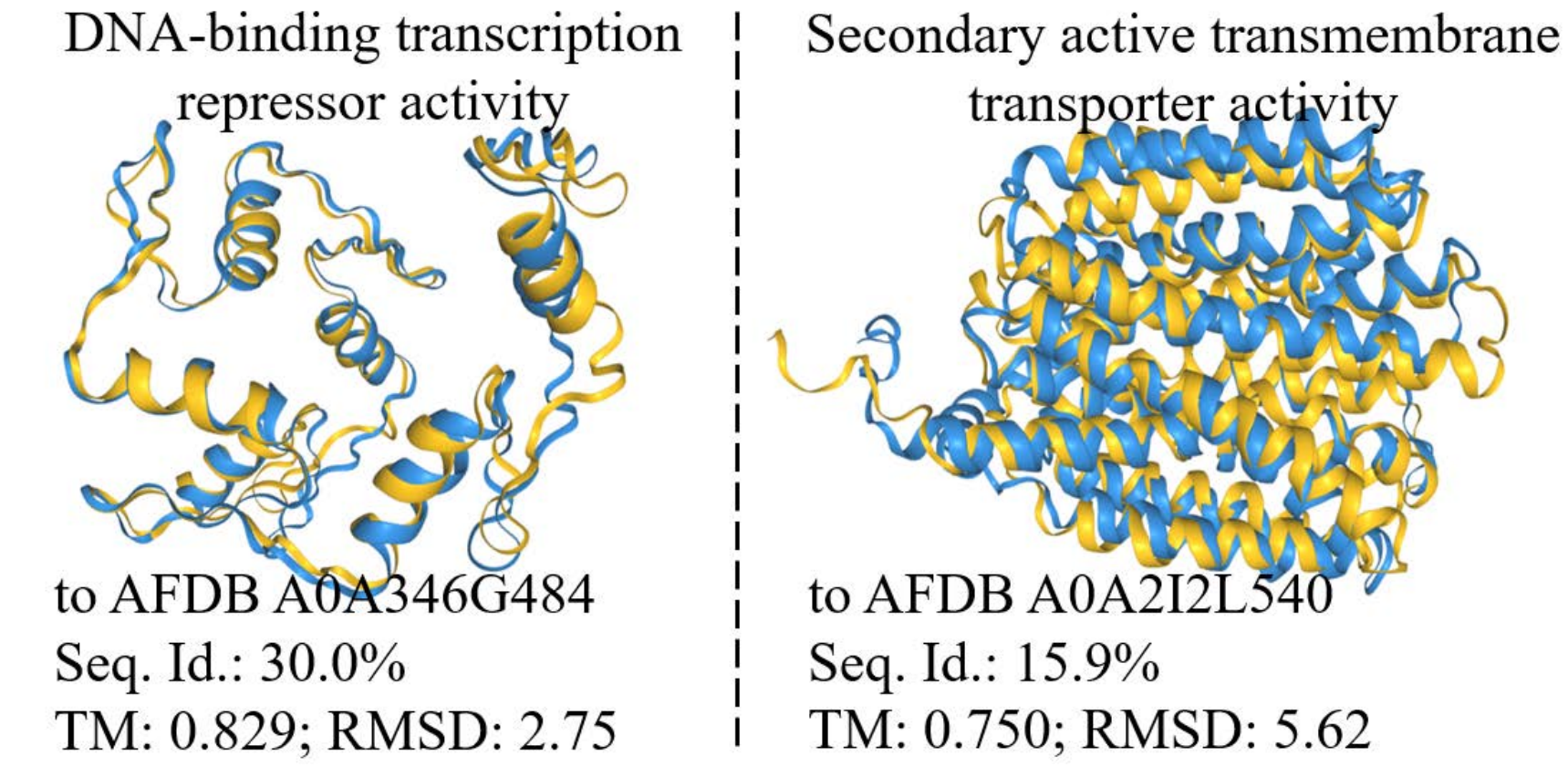}  
    \caption{Intervention case study}
    \label{fig:case_study}
\end{figure}

\smallskip
\noindent\textbf{Case study.}
Figure~\ref{fig:case_study} visualizes proteins generated by \modelname after intervention (in blue).
We identify their most similar natural counterparts (in yellow) with Foldseek~\cite{foldseek}. 
After intervention, \modelname can generate proteins with high structural similarity to natural counterparts with relevant concepts, remaining low sequence identity.
This demonstrates that \modelname effectively captures concept-specific structural features and can successfully steer PLM's generation.
For example, we intervene on the concept of ``\textit{DNA-binding transcription repressor activity}'' and generate a protein structurally similar to the natural protein ``A0A346G484'' (TM-score: 0.829, RMSD:2.75), while maintaining sequence novelty (Seq. ID: 30.0\%).
``A0A346G484'' contains a putative zinc-finger domain and is annotated with the desired concept.
We also generate proteins with high structural similarity to natural proteins exhibiting transmembrane transporter activity.

\section{Conclusion}

We propose \modelname, a semantically-guided SAE to tackle semantic entanglement in SAE training and improve interpretability of PLMs.
We introduce domain knowledge into \modelname to constraint the relationship among concepts, and apply forced activations and feature rescaling to ensure that the learned features effectively contribute to the reconstruction while maintaining high reconstruction fidelity.
Interpretability experiments show that \modelname consistently captures features more aligned with protein structures and functions.
Performance analyses and steering experiments show the superiority of \modelname against existing SAE baselines.

\section*{Acknowledgments}
This work is supported by the National Natural Science Foundation of China (No. 62272219).

\bibliography{aaai2026}



\appendix

\section{Derivation of ELEmbeddings in SAE}
\label{appendix:derivationSAE}

\subsection{Structural Equivalence Between ELEmbeddings Prediction Function and SAE Encoder}
\label{appendix:equivalence}

We formally establish the structural equivalence between the ELEmbeddings prediction function and the forward pass of the encoder in SAE.

In the ELEmbeddings framework, given a protein sequence $p$  and a concept class $c_i$, the prediction score is defined as
\begin{equation}
y'_i = \sigma\left(f_\eta(p)^\top \cdot \left(f_\eta(hF) + f_\eta(c_i)\right) + r_\eta(c_i)\right),
\label{eq:elembeddings_form}
\end{equation}
where $f_\eta(\cdot)$  denotes a projection into the ontology embedding space of dimension $d$, $hF$ represents the hierarchical context of the ontology, and $r_\eta(c_i) > 0$ is a learnable bias term corresponding to a soft radius threshold.
In our setting, protein functions are inferred by performing token-level semantic predictions, where each token representation is evaluated w.r.t. specific biological functions.
The protein $p$ is first tokenized as $p = \{t_1, t_2, \dots, t_L\}$, and each token $t_j$ is embedded using a pretrained language model. 
Let $\mathbf{x_j} \in \mathbb{R}^d$ denote the hidden representation of token $t_j$, which we treat as the input to the encoder. That is, we interpret
\begin{equation}
\label{eq:protein2token}
f_\eta(t_j) = \mathbf{x}_j - \mathbf{b}_{\text{dec}},
\end{equation}
and define
\begin{equation}
\mathbf{w}_i := f_\eta(hF) + f_\eta(c_i) \in \mathbb{R}^d, {b_i} := r_\eta(c_i) \in \mathbb{R}_{>0}.
\end{equation}

Then, Eq.~\eqref{eq:elembeddings_form} becomes, for each token $j$,
\begin{equation}
y'_{i,j} = \sigma\left(\mathbf{w}_i^\top (\mathbf{x}_j - \mathbf{b}_{\text{dec}}) + {b}_i\right),
\label{eq:elembeddings_simplified}
\end{equation}
which mirrors the token-level prediction of a sparse encoder.

In \modelname, each concept $c_i$ is predicted per token as
\begin{equation}
\pi_{\text{pred}}^{(i,j)} = \sigma\left(\mathbf{W}_{\text{pred}}^{(i)} (\mathbf{x_j} - \mathbf{b}_{\text{dec}}) + {b}_{\text{pred}}^{(i)}\right),
\label{eq:sparse_encoder}
\end{equation}
where \( \mathbf{W}_{\text{pred}}^{(i)} \in \mathbb{R}^d \) and \( \mathbf{b}_{\text{pred}}^{(i)} \in \mathbb{R} \) are learned parameters.

By assigning
\begin{align}
\mathbf{W}_{\text{pred}}^{(i)} &:= \mathbf{w}_i^\top = f_\eta(hF) + f_\eta(c_i), \\
{b}_{\text{pred}}^{(i)} &:= {b}_i = r_\eta(c_i), \quad \text{with } {b}_i > 0
\label{eq:equivalence_condition}
\end{align}
we obtain
\begin{equation}
\pi_{\text{pred}}^{(i,j)} = y'_{i,j},
\end{equation}
which demonstrates the structural equivalence of Eq.~\eqref{eq:sparse_encoder} and Eq.~\eqref{eq:elembeddings_form} at the token level.

To compute the sequence-level concept prediction $\pi_{\text{pred}}^{(i)}$ for the full protein sequence, token-level predictions are aggregated (e.g., via max-pooling or average-pooling) across all positions, which is similar to the interpretability procedure in traditional SAE:
\begin{equation}
\pi_{\text{pred}}^{(i)} = \text{Pool}_j\left( \pi_{\text{pred}}^{(i,j)} \right).
\end{equation}

From Eq.~\eqref{eq:equivalence_condition}, we establish a direct correspondence between the encoder weights of the SAE and the terms embedding defined in ELEmbeddings.
Under the constraint $\mathbf{b_i} > 0$, which ensures a valid geometric interpretation of $\mathbf{b_i}$ as the radius of a high-dimensional sphere, the encoder weight vector $\mathbf{W}_{\rm pred}^{(i)}$ learned by the SAE can be interpreted as the sum of two components: the embedding of the ontology concept $c_i $, and the offset vector representing the hasFunction relation $f_\eta(hF)$. 
Therefore, the direction of each encoder unit in the SAE corresponds to the logical composition $ f_\eta(hF) + f_\eta(c_i) $ in the ELEmbeddings framework.

This structural equivalence implies that the ontology-aware semantic geometry encoded by ELEmbeddings is preserved in the SAE encoder.
As a result, the inter-term relations o f the concepts, such as subsumption or regulation,can be explicitly modeled by analyzing the encoder matrix $ \mathbf{W}_{\rm pred} $.
This provides a principled way to ground LLM activations in interpretable, structured biological semantics.

\subsection{Training SAE with ELEmbeddings Axioms}
\label{appendix:axioms}
ELEmbeddings utilize four normalized axiom forms (NF1 to NF4) to encode ontological constraints.
For each normalized form, a specific geometric loss function is defined.
Below, we present the detailed loss formulations and explanations.

From Eq.~\eqref{eq:equivalence_condition}, we observe that the ontology-based representation $f_d(c_i)$  for a given concept $c_i$ can be directly computed using the encoder weight  $\mathbf{W}_{\text{pred}}^{(i)}$ obtained from the SAE. 
This correspondence enables us to directly compute the ELEmbeddings losses with the encoder weights, without requiring separate embedding training.
For all involved relations, we initialize them in the same way as in $\mathbf{W}_{pred}$.

\paragraph{NF1 Loss.}
NF1 corresponds to simple subclass axioms of the form $c_i \sqsubseteq c_j$, e.g., ``\textit{binding}'' (GO:0005488) \texttt{SubClassOf} ``\textit{molecular function}'' (GO:0003674).
The corresponding loss penalizes the distance between the centers of the two $n$-balls and ensures that the $n$-ball for $c_i$ is fully contained in the $n$-ball for $c_j$:
\begin{align}
\mathcal{L}_{\text{NF1}} 
&= \frac{1}{|\text{NF1}|} \sum_{c_i,c_j \in \text{NF1}} 
\max \left(0, \| f_\eta(c_i) - f_\eta(c_j) \| \right. \nonumber \\
&\quad + r_\eta(c_i) - r_\eta(c_j) - \gamma \left. \right) \nonumber \\
&= \frac{1}{|\text{NF1}|} \sum_{i,j \in \text{NF1}} 
\max \left(0, \| \mathbf{w_i} - \mathbf{w_j} \| + {b_i} - {b_j} - \gamma \right).
\label{eq:nf1_loss_reparam}
\end{align}

\paragraph{NF2 Loss.}
NF2 handles axioms of the form $c_i \sqcap c_j \sqsubseteq c_k$, which express that the intersection of two concepts is a subclass of a third, e.g., ``\textit{cutinase activity}'' (GO:0050525) and ``\textit{biological regulation}'' (GO:0065007) \texttt{SubClassOf} ``\textit{positive regulation of protein kinase B signaling}'' (GO:0051897).
The corresponding loss minimizes the discrepancy between the intersection of $n$-balls for $c_i$ and $c_j$, and the $n$-ball for class $E$:

\begin{align}
\mathcal{L}_{\text{NF2}}  &= \frac{1}{|\text{NF2}|} \sum_{c_i,c_j,c_k \in \text{NF2}} \Big(
\max(0, \| f_\eta(c_i) - f_\eta(c_j) \|  \nonumber\\
&\quad - r_\eta(c_i) - r_\eta(c_j) - \gamma) \nonumber \\
&\quad + \max(0, \| f_\eta(c_i) - f_\eta(c_k) \| - r_\eta(c_i) - \gamma) \nonumber\\
&\quad + \max(0, \| f_\eta(c_j) - f_\eta(c_k) \| - r_\eta(c_j) - \gamma) \nonumber\\
&\quad + \max(0, \min(r_\eta(c_i), r_\eta(c_j)) - r_\eta(c_k) - \gamma) \Big) \nonumber\\
&= \frac{1}{|\text{NF2}|} \sum_{i,j,k \in \text{NF2}} \Big(
\max(0, \| \mathbf{w_i} - \mathbf{w_j} \| - {b_i} - {b_j} - \gamma) \nonumber\\
&\quad + \max(0, \| \mathbf{w_i} - \mathbf{w_k} \| - {b_i} - \gamma) \nonumber\\
&\quad + \max(0, \| \mathbf{w_j} - \mathbf{w_k} \| - {b_j} - \gamma) \nonumber\\
&\quad + \max(0, \min({b_i}, {b_j}) - {b_k} - \gamma) \Big). 
\end{align}

\paragraph{NF3 Loss.}
NF3 corresponds to axioms of the form $c_i \sqsubseteq \exists R.c_j$, indicating that concept $c_i$ is included in the set of entities that are related by $R$ to some instance of $c_j$, e.g., ``\textit{positive regulation of arginine biosynthetic process}'' (GO:1900080) \texttt{SubClassOf} ``\textit{positively regulates}'' (RO:0002213) some ``\textit{arginine biosynthetic process}'' (GO:0006526).
This is modeled by translating the $n$-ball of class $c_j$ by the relation vector $f_\eta(R)$ and minimizing its non-overlap with $c_i$:

\begin{align}
\mathcal{L}_{\text{NF3}} &= \frac{1}{|\text{NF3}|}\sum_{R,c_i,c_j \in \text{NF3}} 
\max \left(0, \| f_\eta(c_i) - f_\eta(R) - f_\eta(c_j) \| \right. \nonumber\\
&\quad - r_\eta(c_i) - r_\eta(c_j) - \gamma \left. \right) \nonumber\\
&= \frac{1}{|\text{NF3}|} \sum_{R,c_i,c_j \in \text{NF3}}
\max \left(0, \| \mathbf{w_i}  - \mathbf{w_j} - f_\eta(R) \| \right. \nonumber\\
&\quad- {b_i} - {b_j} - \gamma \left. \right).
\end{align}

\paragraph{NF4 Loss.}
NF4 axioms are of the form $\exists R.c_i \sqsubseteq c_j$, implying that entities related by $R$ to some instance of $c_i$ are contained in class $c_j$, e.g., \textit{part of} (BFO:0000050) some ``\textit{conjugation}'' (GO:0000746) \texttt{SubClassOf} ``\textit{mammary stem cell proliferation}'' (GO:0002174).
The loss translates the $n$-ball of $c_i$ by relation vector $f_\eta(R)$, and ensures containment within $c_j$'s $n$-ball:

\begin{align}
\mathcal{L}_{\text{NF4}} &= \frac{1}{|\text{NF4}|} \sum_{c_i,R,c_j \in \text{NF4}} 
\max \left(0, \| f_\eta(c_i) + f_\eta(R) - f_\eta(c_j) \| \right. \nonumber\\
&\quad+ r_\eta(c_i) - r_\eta(c_j) - \gamma \left. \right) \nonumber\\
&= \frac{1}{|\text{NF4}|} \sum_{c_i,R,c_j \in \text{NF4}} 
\max \left(0, \| \mathbf{w_i} - \mathbf{w_j} + f_\eta(R)\| \right. \nonumber\\
&\quad+ {b_i} - {b_j} - \gamma \left. \right).
\end{align}

\section{Pseudocode}
\label{appendix:pseudocode}

Please see Algorithms~\ref{alg:forward} and \ref{alg:training} below.

\begin{algorithm}[!ht]
\caption{Forward Pass with SAE}
\label{alg:forward}
\KwIn{Input $\mathbf{x} \in \mathbb{R}^d$}
\KwOut{Reconstruction $\mathbf{\hat{x}}$, semantic prediction $\pi_{\rm pred}$}

$\mathbf{W}_{\rm def} \leftarrow \mathbf{W}_{\rm pred}^{\rm detach} \cdot \exp(\mathbf{r}_{\rm pred})$ \\
\tcc{\small Eq.~\eqref{eq:weight_tying}, compute weight matrix}

$\mathbf{z}_{\rm unk} \leftarrow \mathrm{TopK}(\mathbf{W}_{\rm unk}(\mathbf{x} - \mathbf{b}_{\rm dec}) + \mathbf{b}_{\rm unk})$\\
\tcc{\small Eq.~\eqref{eq:encoding}, compute activations}
$\mathbf{\hat{z}}_{\rm def} \leftarrow \mathbf{W}_{\rm def}(\mathbf{x} - \mathbf{b}_{\rm dec}) + \mathbf{b}_{\rm def}$ 

$\mathbf{z}_{\rm bias} \leftarrow \mathds{1}_{\pi_{\rm pred} > 0} \cdot \mathrm{ReLU}(\mathrm{mean}(\mathbf{z}_{\rm unk}) - \mathbf{\hat{z}}_{\rm def})$ \\
\tcc{\small Eq.~\eqref{eq:bias}, force activation}

$\mathbf{z}_{\rm def} \leftarrow \mathbf{\hat{z}}_{\rm def} + \mathbf{z}_{\rm bias}$ \\
\tcc{\small Eq.~\eqref{eq:bias}, defined activations}

$\mathbf{z} \leftarrow \texttt{Concat}(\mathbf{z}_{\rm def}, \mathbf{z}_{\rm unk})$

$\mathbf{\hat{x}} \leftarrow \mathbf{W}_{\rm dec} \mathbf{z} + \mathbf{b}_{\rm dec}$ \\
\tcc{\small Eq.~\eqref{eq:reconstruction}, reconstruction}

$\pi_{\rm pred} \leftarrow \sigma(\mathbf{W}_{\rm pred}(\mathbf{x} - \mathbf{b}_{\rm dec}) + \mathbf{b}_{\rm pred})$\\
\tcc{\small Eq.~\eqref{eq:prediction}, prediction}

\Return{$\mathbf{\hat{x}}$, $\pi_{\rm pred}$}
\end{algorithm}

\begin{algorithm}[!ht]
\caption{Training Procedure}
\label{alg:training}
\KwIn{Dataset $\mathcal{D} = \{(\mathbf{x^{(j)}}, y^{(j)})\}$ with semantic labels $y^{(j)} \in \{0,1\}^m$ }

\ForEach{mini-batch $\{\mathbf{x}, y\} \subset \mathcal{D}$}{

    Compute $\mathbf{z}_{\rm def}, \mathbf{z}_{\rm unk}, \pi_{\rm pred}, \mathbf{\hat{x}}$ using Algorithm~\ref{alg:forward}
    
    $\mathcal{L}_{\rm rec} \leftarrow \|\mathbf{x} - \mathbf{\hat{x}}\|_2^2$ \\
    \tcc{\small Reconstruction loss}
    
    $\mathcal{L}_{\rm annot} \leftarrow \mathrm{CrossEntropy}(\pi_{\rm pred}, y)$ \\
    \tcc{\small Annotation guidance loss}
    $\mathcal{L}_{\rm axioms} \leftarrow \mathcal{L}_{\rm NF1} + \mathcal{L}_{\rm NF2} + \mathcal{L}_{\rm NF3} + \mathcal{L}_{\rm NF4}$ \\
    \tcc{\small Get normalized axiom loss}
    
    $\mathcal{L} \leftarrow \mathcal{L}_{\rm rec} + \lambda_{\rm annot} \mathcal{L}_{\rm annot} + \lambda_{\rm axiom} \mathcal{L}_{\rm axiom}$\\
    \tcc{\small Total loss}
    
    Backpropagate $\mathcal{L}$ and update all trainable parameters
}
\end{algorithm}

\section{Interpretability Experiments Appendix}
\label{appendix:interpretability_experiments_setting}

\subsection{Settings}
To better demonstrate the superior interpretability of \modelname, we design comprehensive interpretability experiments, including both relevance-based and probing-based evaluations.
Furthermore, we present several cases on \modelname interpretability including protein functional structure and metal ion binding site prediction.

For probing-based interpretation, we perform experiments on protein function prediction tasks.
This task aims to uncover the biological roles and interactions of proteins, which is critical for applications such as drug target identification, understanding disease mechanisms, and advancing biotechnology.
Protein annotations are sourced from the Gene Ontology (GO), which is structured into three sub-ontologies: Molecular Function (MFO), Biological Process (BPO), and Cellular Component (CCO).

For relevance-based interpretation, we analyze 15 biologically meaningful concepts drawn from three distinct datasets.
For each concept, we retrieve relevant proteins from UniProtKB, ensuring that all training samples are excluded to construct an independent evaluation set consisting of 5,000 proteins.
In this analysis, the top-10 most activated features identified by the model are selected as candidate features for interpretation.

\subsection{Datasets}
We adopt the protein function prediction benchmark from the previous work~\cite{deepgo2}.\footnote{The dataset is publicly available under the BSD 3-Clause License.}
The dataset is extracted from UniProtKB/Swiss-Prot and is filtered to retain those with experimental annotations supported by evidence codes such as EXP, IDA, IPI, IMP, IGI, IEP, TAS, IC, HTP, HDA, HMP, HGI, and HEP.
This filtering resulted in a high-quality dataset containing 77,647 manually curated proteins.
The Gene Ontology release used corresponds to November 16, 2021.
Models are trained and evaluated separately for each GO sub-ontology.
Detailed information is provided in Table~\ref{tab:dataset_info}. 
It shows the number of GO terms, total number of proteins, number of groups of similar proteins, number of proteins in training, validation and testing sets.

\begin{table*}[t]
\centering
{\small
\begin{tabular}{lrrrrrr}
\toprule
Ontology & GO Terms & Proteins & Groups & Training & Validation & Testing \\
\midrule
MFO & 6,851  & 43,279 & 6,963  & 52,072 & 2,964  & 4,221 \\
BPO & 21,356 & 58,729 & 9,463  & 52,584 & 2,870  & 3,275 \\
CCO & 2,829  & 59,257 & 10,019 & 48,318 & 4,970  & 5,969 \\
\bottomrule
\end{tabular}}
\caption{Summary of the UniProtKB/Swiss-Prot dataset}
\label{tab:dataset_info}
\end{table*}

To further assess the interpretability of \modelname, we conduct experiments on the metal ion-binding datasets~\cite{metal_ion_binding}.  
These datasets cover four biologically relevant ion types: Zn$^{2+}$, Ca$^{2+}$, Mg$^{2+}$, and Mn$^{2+}$.  
Detailed statistics for each subset are provided in Table~\ref{tab:metal_binding_stats}.

\begin{table*}[t]
\centering
{\small
\begin{tabular}{clrr}
\toprule
Ligand type & Dataset & Binding residue & Non-binding residue \\
\midrule
\multirow{2}{*}{Zn$^{2+}$} 
    & ZN\_Train\_1647 & 7,731 & 467,184 \\
    & ZN\_Test\_211   & 1,039 & 54,981  \\
\midrule
\multirow{2}{*}{Ca$^{2+}$} 
    & CA\_Train\_1554 & 8,442 & 495,700 \\
    & CA\_Test\_183   & 1,034 & 65,820  \\
\midrule
\multirow{2}{*}{Mg$^{2+}$} 
    & MG\_Train\_1730 & 6,321 & 569,572 \\
    & MG\_Test\_235   &  893 & 87,913  \\
\midrule
\multirow{2}{*}{Mn$^{2+}$} 
    & MN\_Train\_547  & 2,556 & 179,143 \\
    & MN\_Test\_57    &  225 & 20,194  \\
\bottomrule
\end{tabular}}
\caption{Statistics of the metal ion binding-sites prediction dataset}
\label{tab:metal_binding_stats}
\end{table*}

\subsection{Implementations}
We train all SAE models on protein function prediction datasets, using a learning rate of 5e-4 and a batch size of 12,800 for 25,000 steps. For \modelname and TopK SAE, we vary the number of active neurons with $K \in \{50, 100, 500, 1000\}$. 
For Gated SAE, we tune the L1 regularization coefficient in $\{1.5\text{e-}4,$ $2\text{e-}4, 3\text{e-}4, 4\text{e-}4, 5\text{e-}4\}$.
For Naive SAE, we explore L1 coefficients in $\{8\text{e-}5, 6\text{e-}5, 2\text{e-}4, 3\text{e-}4, 4\text{e-}4\}$.
We set the activation width to 40,000 for the BPO dataset and 30,000 for both MFO and CCO.
The dimensionality of the activation vectors matches that of the ESM2-15B model's hidden states (5,120).
We fix both $\lambda_{\text{sup}}$ and $\lambda_{\text{axiom}}$ to 1.
All results are reported using the representations from layer 35 of ESM2-15B. 
We train \modelname on the metal-ion binding dataset using a learning rate of $5 \times 10^{-4}$ and a batch size of 3200 for 40{,}000 optimization steps.  
The activation dimension of \modelname is set to 10{,}000.  
For evaluation, we compute top-$K$ activations with $K \in \{50,\ 100,\ 500,\ 1000\}$ to analyze feature sparsity and relevance. 
All experiments are conducted on four NVIDIA A800 GPUs.

\subsection{Baselines and Metrics}
\label{appendix:baseline_metric}
We compare \modelname with several representative SAE baselines, a dictionary learning method SpLiCE~\cite{splice} and linear probe on PLMs hidden representations:
\begin{itemize}
    \item \textbf{Naive SAE:} A standard SAE trained only with reconstruction loss and a simple L1 penalty for sparsity.
    \item \textbf{Gated SAE:} An enhanced SAE variant that applies learned gating mechanisms and L1 regularization to encourage selective activation.
    \item \textbf{TopK SAE:} A Top-K SAE that enforces hard sparsity by retaining only the top-K highest activations for each input.
    \item \textbf{SpLiCE:} A dictionary learning method that transforms representations into sparse linear combinations of human interpretable concepts.
    
    \item \textbf{Linear Probe:} We direct use linear probe on LLMs' internal representation, to show the infromation loss of SAE methods.
\end{itemize}

\paragraph{Metrics.}
For relevance-based interpretation, 
following \cite{interplm, plmsae}, we use the F1 score to evaluate the relevance of selected features.
The detailed calculation is as follows:
\begin{align}
\text{precision} &= \frac{\text{TruePositives}}{\text{TruePositives} + \text{FalsePositives}}, \\
\text{recall} &= \frac{\text{TruePositive}}{\text{TruePositives}+\text{FalseNegative}}, \\
\text{F1} &= 2 \cdot \frac{\text{precision} \cdot \text{recall}}{\text{precision} + \text{recall}}.
\end{align}

For each feature, we determine whether it is positive based on whether its normalized activation value exceeded a specific threshold. For each concept, we searched thresholds from the set \{0, 0.05, 0.1, 0.15, 0.2, 0.25, 0.5, 0.8\}, and selected the threshold that yielded the highest F1 score.

For probing-based prediction, we adopt AUPR and a class-centric AUC from~\cite{deepgozero,deepgo2}.
We use the protein-centric evaluation metrics $F_{\max}$ and $S_{\min}$ introduced by the CAFA challenge~\cite{fmax_1, fmax_2}.
The detailed formulations of metrics are provided below.

$F_{\text{max}}$ is a maximum protein-centric F-measure computed over all prediction thresholds.
First, it computes average precision and recall using the following equations:
\begin{align}
    pr_u(\tau) &= \frac{\sum_c I(c \in \hat{T}_u(\tau) \land c \in T_u)}{\sum_c I(c \in \hat{T}_u(\tau))},\\
    rc_u(\tau) &= \frac{\sum_c I(c \in \hat{T}_u(\tau) \land c \in T_u)}{\sum_c I(c \in T_u)},\\
    \text{AvgPr}(\tau) &= \frac{1}{|\texttt{set}(\tau)|} \sum_{u \in \texttt{set}(\tau)} pr_u(\tau),\\
    \text{AvgRc}(\tau) &= \frac{1}{N} \sum_{u=1}^{N} rc_u(\tau),
\end{align}
where $c$ is a GO class, $T_u$ is the set of true annotations, $\hat{T}_u(\tau)$ is the set of predicted annotations for protein $u$ at threshold $\tau$, $\texttt{set}(\tau)$ is the set of proteins for which at least one GO class is predicted at threshold $\tau$, $N$ is the total number of proteins, and $I$ is the indicator function returning 1 if the condition holds and 0 otherwise. Then, we compute $F_{\text{max}}$ over thresholds $\tau \in [0, 1]$ with a step size of 0.01. A class $c$ is considered predicted for protein $u$ if its score is greater than or equal to $\tau$:
\begin{equation}
    F_{\text{max}} = \max_\tau \left\{ \frac{2 \cdot \text{AvgPr}(\tau) \cdot \text{AvgRc}(\tau)}{\text{AvgPr}(\tau) + \text{AvgRc}(\tau)} \right\}.
\end{equation}

$S_{\text{min}}$ computes the semantic distance between real and predicted annotations based on information content of the classes. The information content $IC(c)$ is computed based on the annotation probability of class $c$:
\begin{equation}
    IC(c) = -\log (Pr(c \mid \mathcal{P}(c))),
\end{equation}
where $\mathcal{P}(c)$ denotes the parent classes of $c$. The $S_{\text{min}}$ is computed using the following equation:
\begin{equation}
    S_{\text{min}} = \min_\tau \sqrt{ru(\tau)^2 + mi(\tau)^2},
\end{equation}
where $ru(\tau)$ is the average remaining uncertainty and $mi(\tau)$ is the average misinformation:
\begin{align}
    ru(\tau) &= \frac{1}{N} \sum_{u=1}^{N} \sum_{c \in T_u \setminus \hat{T}_u(\tau)} IC(c), \\
    mi(\tau) &= \frac{1}{N} \sum_{u=1}^{N} \sum_{c \in \hat{T}_u(\tau) \setminus T_u} IC(c).
\end{align}

\subsection{Metal Ion Binding-Sites Prediction Cases}
\begin{figure}[h]
    \centering
    \begin{subfigure}[b]{0.24\textwidth}
        \centering
        \includegraphics[width=\linewidth]{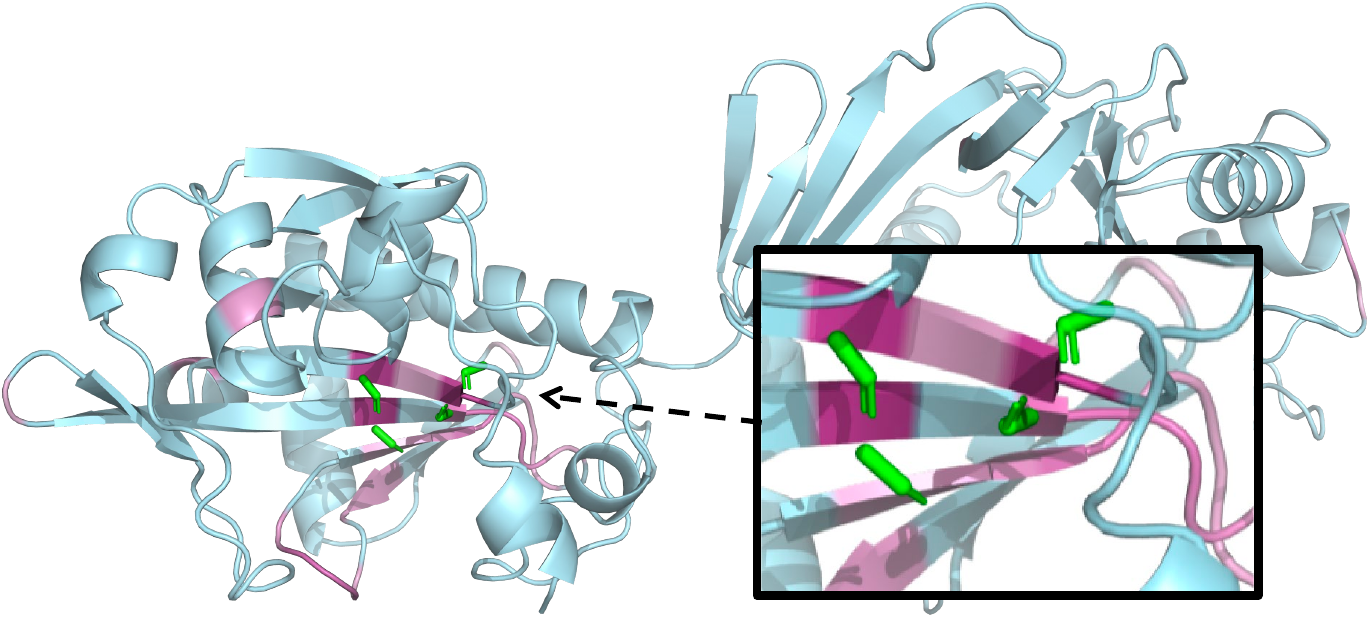}
        \caption{Mg$^{2+}$ binding sites}
        \label{appendix_fig:mg_binding_site}
    \end{subfigure}
    \begin{subfigure}[b]{0.22\textwidth}
        \centering
        \includegraphics[width=\linewidth]{./figures/binding_site_prediction/zn_figure.pdf}
        \caption{Zn$^{2+}$ binding sites}
        \label{appendix_fig:zn_binding_site}
    \end{subfigure}
    \begin{subfigure}[b]{0.24\textwidth}
        \centering
        \includegraphics[width=\linewidth]{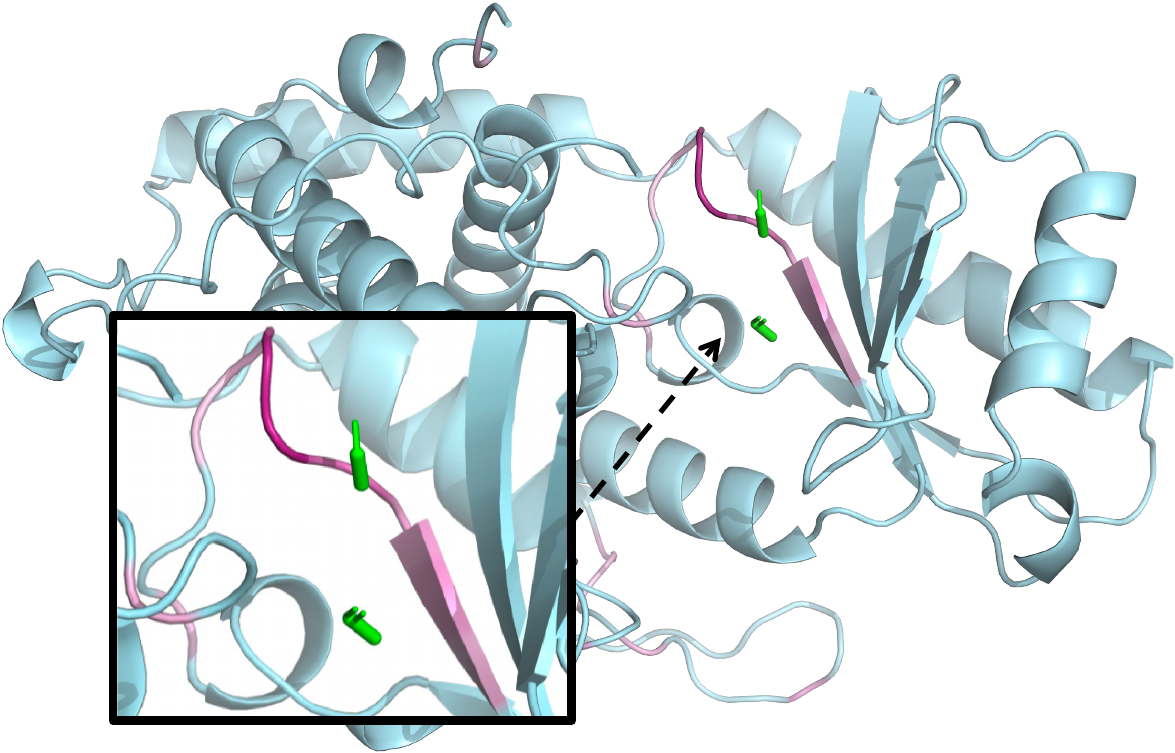}
        \caption{Ca$^{2+}$ binding sites}
        \label{appendix_fig:ca_binding_site}
    \end{subfigure}
    \begin{subfigure}[b]{0.22\textwidth}
        \centering
        \includegraphics[width=\linewidth]{./figures/binding_site_prediction/mn_figure.pdf}
        \caption{Mn$^{2+}$ binding sites}
        \label{appendix_fig:mn_binding_site}
    \end{subfigure}
    \caption{Interpretability results on metal ion binding-sites prediction}
    \label{appendix_fig:ion_binding_result}
\end{figure}

In this section, we present a case study demonstrating the interpretability of ion binding-sites concepts.  
We use the normalized activation score of the feature corresponding to a specific ion binding-sites concept as the prediction signal from \modelname for identifying whether an amino acid residue is part of an ion binding site.  
In Figure~\ref{appendix_fig:ion_binding_result}, regions with higher activation scores are shown in red, while blue spheres indicate the locations of ground-truth binding sites.  
As observed, the learned feature is highly activated at the true binding sites while remaining largely inactive in non-target regions, suggesting a strong correlation between the feature and the intended concept.

\subsection{Relevance-based Interpretation Detailed Results}
\label{appendix:analytic result}
\begin{figure}[t]
    \centering
    \begin{subfigure}[t]{0.49\linewidth}
        \centering
        \includegraphics[width=\linewidth]{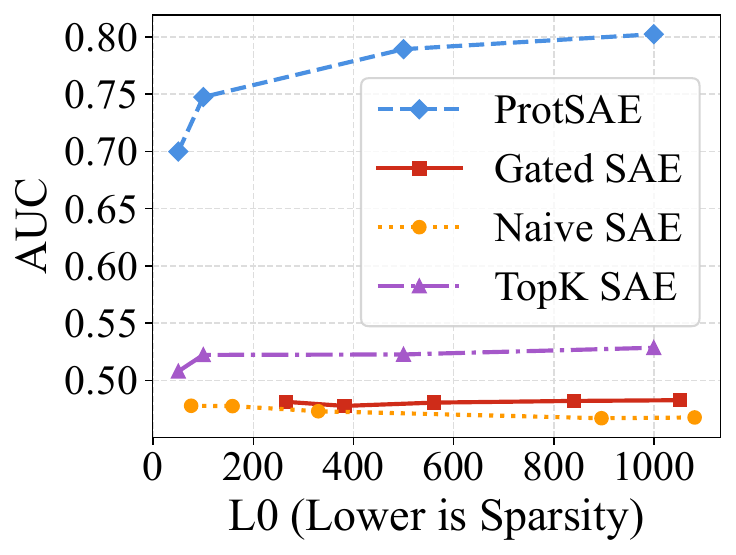}
        \caption{AUC w.r.t. $L_0$}
        \label{fig:mf_auc}
    \end{subfigure}
    \begin{subfigure}[t]{0.49\linewidth}
        \centering
        \includegraphics[width=\linewidth]{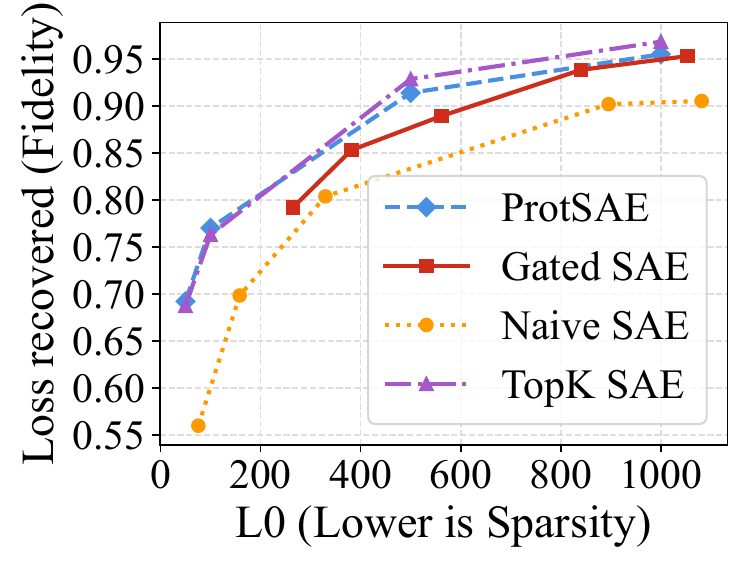}
        \caption{Loss Recovered w.r.t.  $L_0$}
        \label{fig:mf_recovered}
    \end{subfigure}

    \begin{subfigure}[t]{0.49\linewidth}
        \centering
        \includegraphics[width=\linewidth]{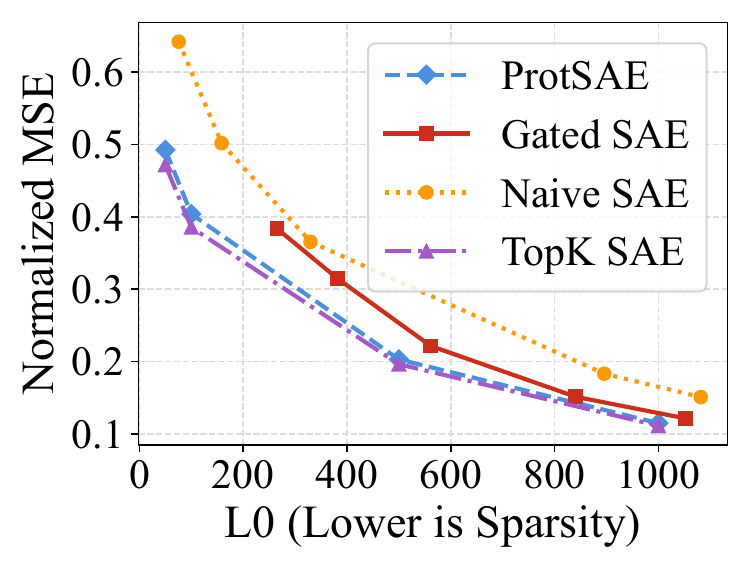}
        \caption{MSE w.r.t. $L_0$}
        \label{fig:mf_mse}
    \end{subfigure}
    \begin{subfigure}[t]{0.49\linewidth}
        \centering
        \includegraphics[width=\linewidth]{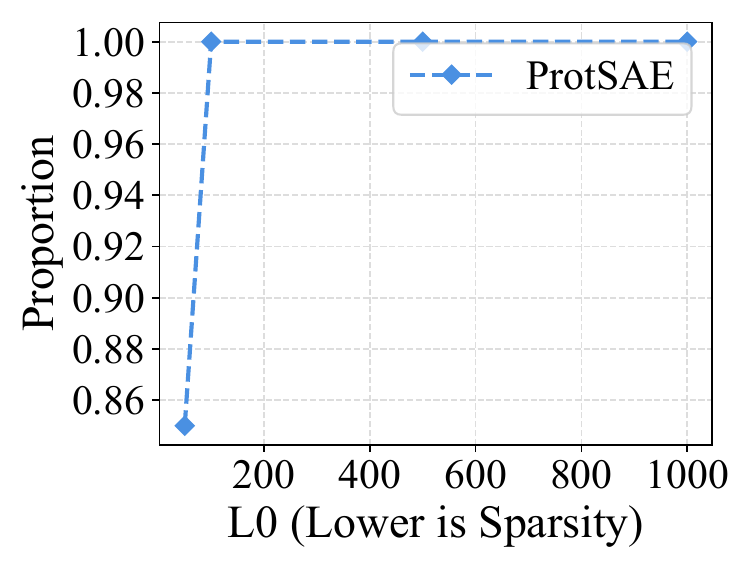}
        \caption{Reconstruction proportion of predicted activations w.r.t. $L_0$}
        \label{fig:mf_proportion}
    \end{subfigure}

    \caption{Performance comparison under different sparsity on the Molecular Function Ontology  dataset}
    \label{fig:mf_group}
\end{figure}

\begin{figure}[t]
    \centering
    \begin{subfigure}[t]{0.49\linewidth}
        \centering
        \includegraphics[width=\linewidth]{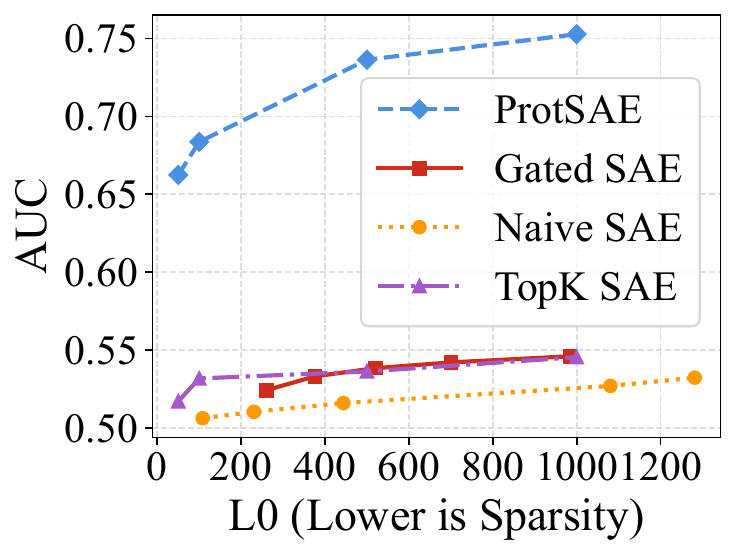}
        \caption{AUC w.r.t. $L_0$}
        \label{fig:bp_auc}
    \end{subfigure}
    \begin{subfigure}[t]{0.49\linewidth}
        \centering
        \includegraphics[width=\linewidth]{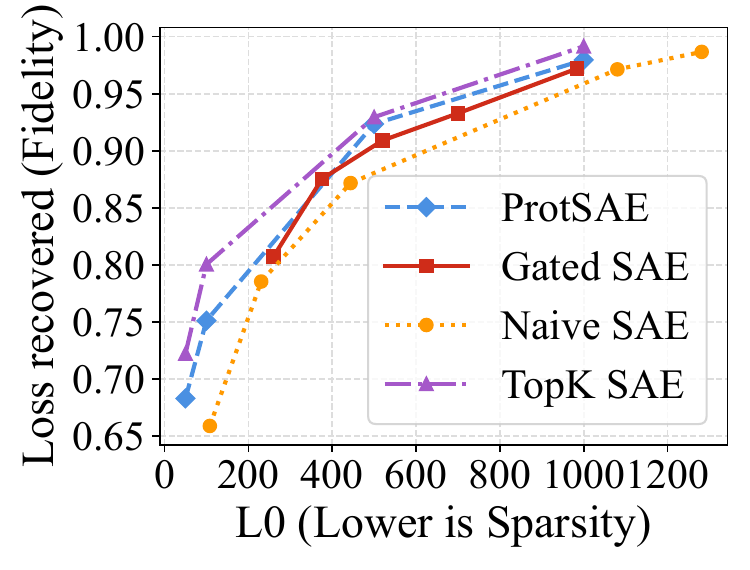}
        \caption{Loss Recovered w.r.t. $L_0$}
        \label{fig:bp_recovered}
    \end{subfigure}

    \begin{subfigure}[t]{0.49\linewidth}
        \centering
        \includegraphics[width=\linewidth]{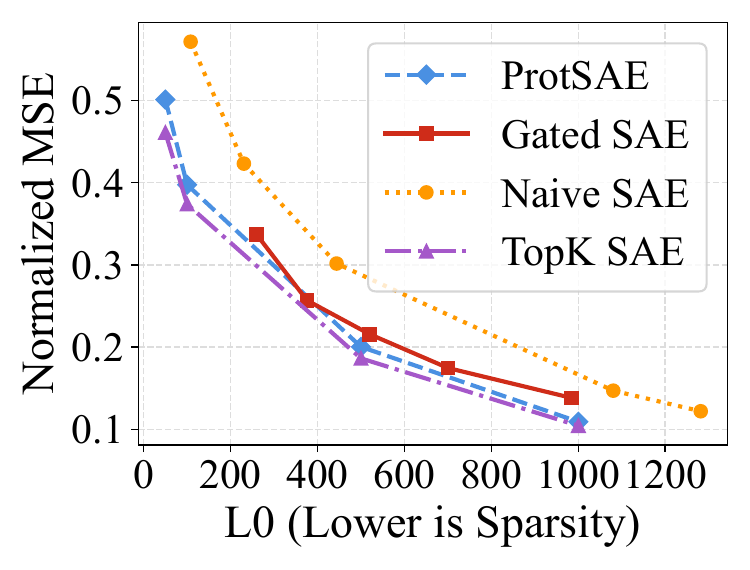}
        \caption{MSE w.r.t. $L_0$}
        \label{fig:bp_mse}
    \end{subfigure}
    \begin{subfigure}[t]{0.49\linewidth}
        \centering
        \includegraphics[width=\linewidth]{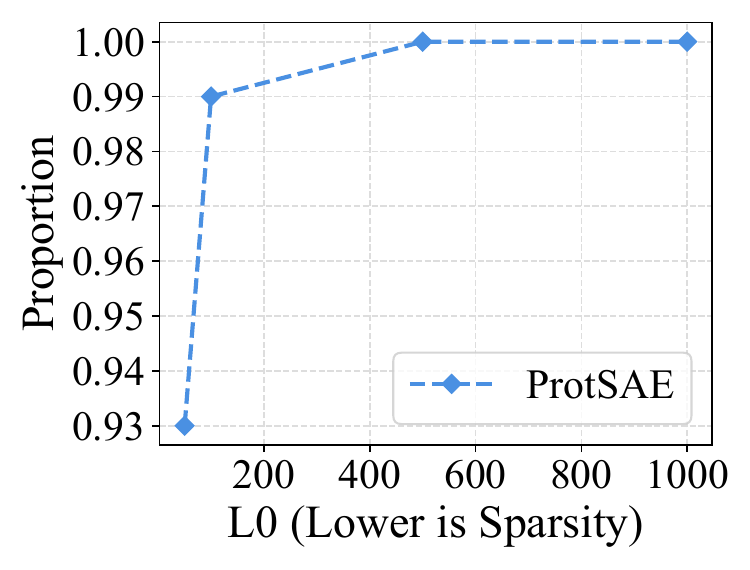}
        \caption{Reconstruction proportion of predicted activations w.r.t. $L_0$}
        \label{fig:bp_proportion}
    \end{subfigure}

    \caption{Performance comparison under different sparsity on the Biological Process Ontology  dataset}
    \label{fig:bp_group}
\end{figure}

\begin{figure}[t]
    \centering
    \begin{subfigure}[t]{0.49\linewidth}
        \centering
        \includegraphics[width=\linewidth]{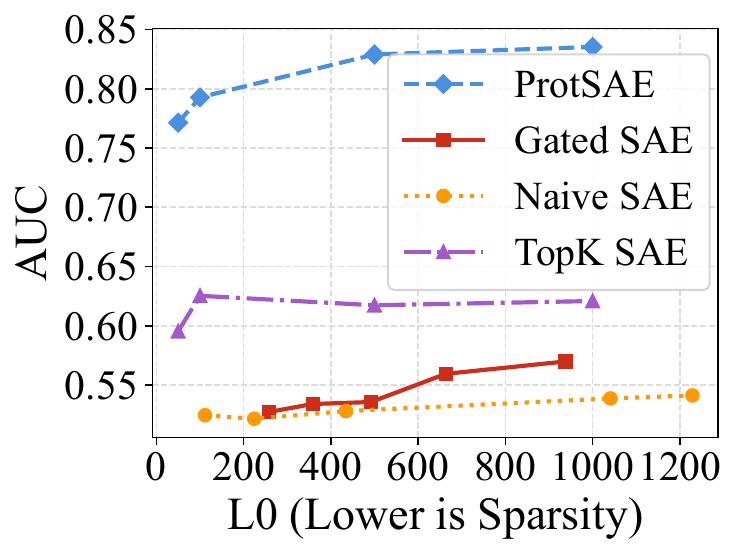}
        \caption{AUC w.r.t. $L_0$}
        \label{fig:cc_auc}
    \end{subfigure}
    \begin{subfigure}[t]{0.49\linewidth}
        \centering
        \includegraphics[width=\linewidth]{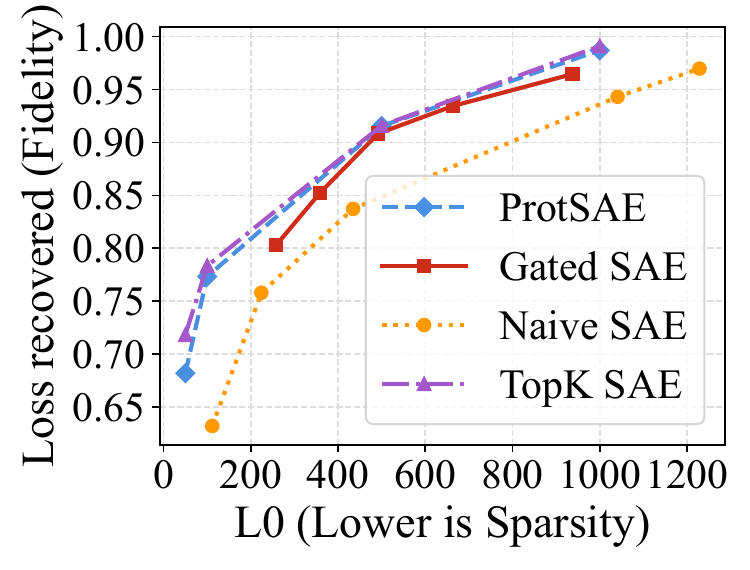}
        \caption{Loss Recovered w.r.t. $L_0$}
        \label{fig:cc_recovered}
    \end{subfigure}

    \begin{subfigure}[t]{0.49\linewidth}
        \centering
        \includegraphics[width=\linewidth]{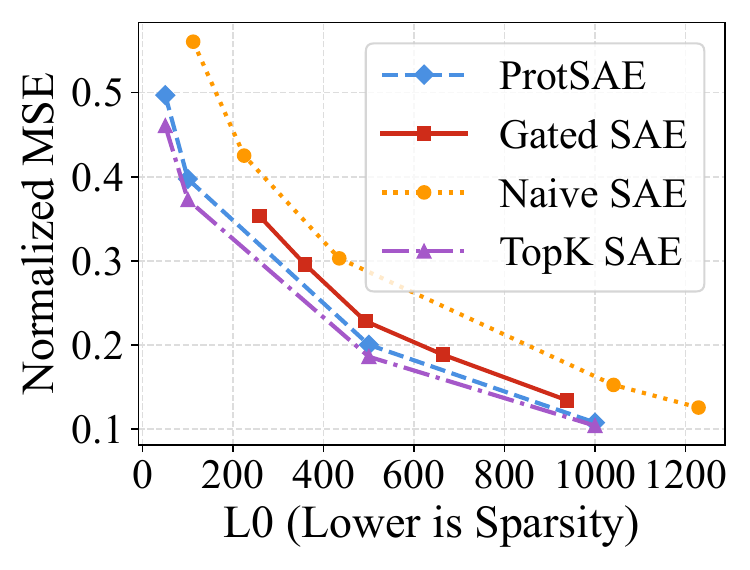}
        \caption{MSE w.r.t. $L_0$}
        \label{fig:cc_mse}
    \end{subfigure}
    \begin{subfigure}[t]{0.49\linewidth}
        \centering
        \includegraphics[width=\linewidth]{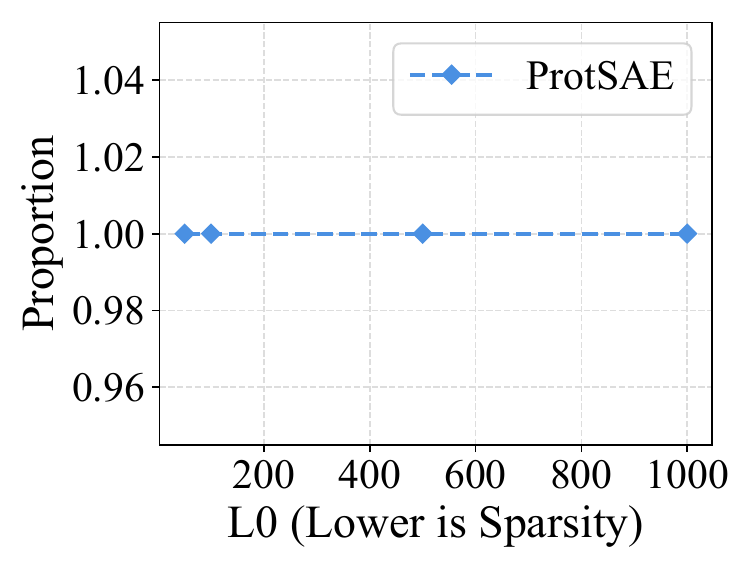}
        \caption{Reconstruction proportion of predicted activations w.r.t. $L_0$}
        \label{fig:cc_proportion}
    \end{subfigure}

    \caption{Performance comparison under different sparsity on the Cellular Component Ontology dataset}
    \label{fig:cc_group}
\end{figure}

We present detailed experimental results on three datasets in Figures~\ref{fig:mf_group}, \ref{fig:bp_group}, and \ref{fig:cc_group}, analyzing the changes in AUC, Loss Recovered, Normalized MSE, and Proportion under different sparsity levels.

\section{Performance Analyses}
\label{appendix:performance_analyses}

\subsection{Metrics}
For evaluating the SAE performance, following~\cite{gatedsae,topksae}, we employ Normalized MSE and Loss Recovered to assess the reconstruction fidelity of the model, and $L_0$ to assess the sparsity.
The formal definitions of these metrics are given below.

The $L_0$ of a SAE is defined by the average number of active features on a given input $\mathbb{E}_{x \sim \mathcal{D}} \left\| f(x) \right\|_0$.

The loss recovered of an SAE is calculated from the average cross-entropy loss of the language model on an evaluation dataset, when the SAE's reconstructions are spliced into it.  
If we denote by $\mathrm{CE}(\phi)$ the average loss of the language model when we splice in a function $\phi : \mathbb{R}^n \rightarrow \mathbb{R}^n$ at the SAE's site during the model’s forward pass, then loss recovered is
\begin{equation}
1 - \frac{\mathrm{CE}(\hat{x} \circ f) - \mathrm{CE}(\mathrm{Id})}{\mathrm{CE}(\zeta) - \mathrm{CE}(\mathrm{Id})},
\end{equation}
where $\hat{x} \circ f$ is the autoencoder function, $\zeta : \mathbf{x} \mapsto \mathbf{0}$ the zero-ablation function, and $\mathrm{Id} : \mathbf{x} \mapsto \mathbf{x}$ the identity function.

We compute the mean squared error MSE between reconstructed output $\hat{x}$ and the ground-truth $x$, normalized by the MSE obtained using the mean $\bar{x}$ as reconstruction:
\begin{equation}
    \text{Normalized\_MSE}(\hat{x}, x) = \frac{\text{MSE}(\hat{x}, x)}{\text{MSE}(\bar{x}, x)}.
\end{equation}

\subsection{Detailed Results}
\label{appendix:protein_function_prediction}
We present experimental results on three protein function prediction datasets in Figures~\ref{fig:mf_group}, \ref{fig:bp_group}, and \ref{fig:cc_group}, analyzing the changes in AUC, Loss Recovered, Normalized MSE, and Proportion under different sparsity levels.
\modelname consistently outperforms baselines in AUC under different sparsity levels, showing strong semantic retention.
Meanwhile, it maintains competitive reconstruction quality, confirming its ability to capture meaningful concepts without losing essential concept information.
Furthermore, the higher reconstruction proportion of predicted activations with respect to $L_0$ indicates that the defined neurons effectively contribute to the reconstruction process.

\begin{figure*}
    \centering

    \begin{subfigure}[t]{0.21\textwidth}
        \centering
        \includegraphics[width=\linewidth]{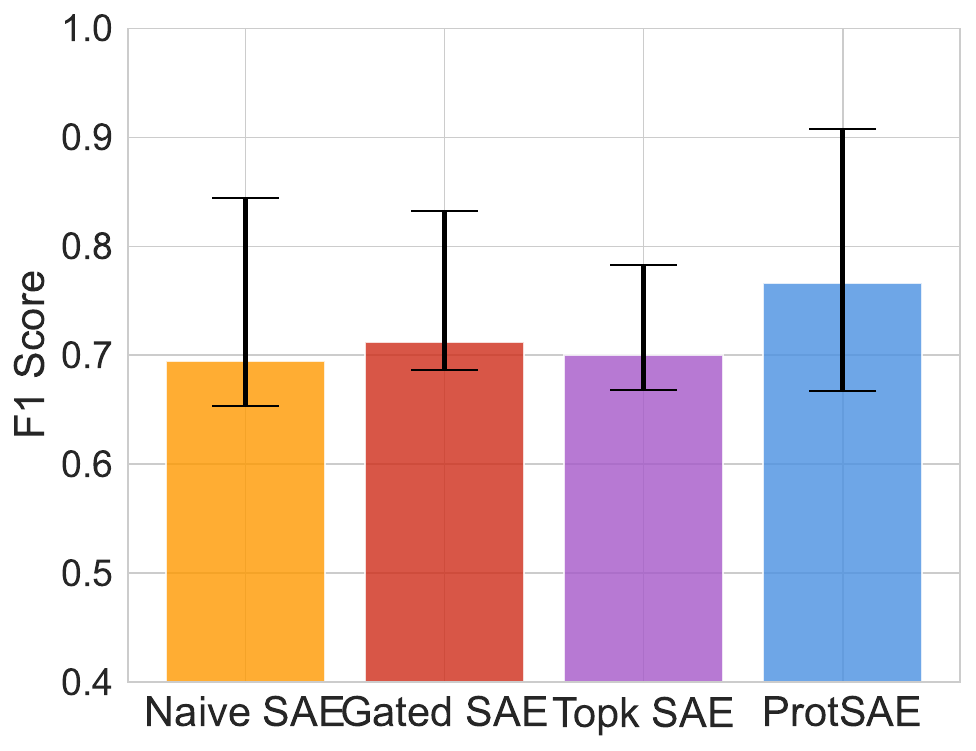}
        \caption{Chloroplast stroma}
    \end{subfigure}
    \hfill
    \begin{subfigure}[t]{0.21\textwidth}
        \centering
        \includegraphics[width=\linewidth]{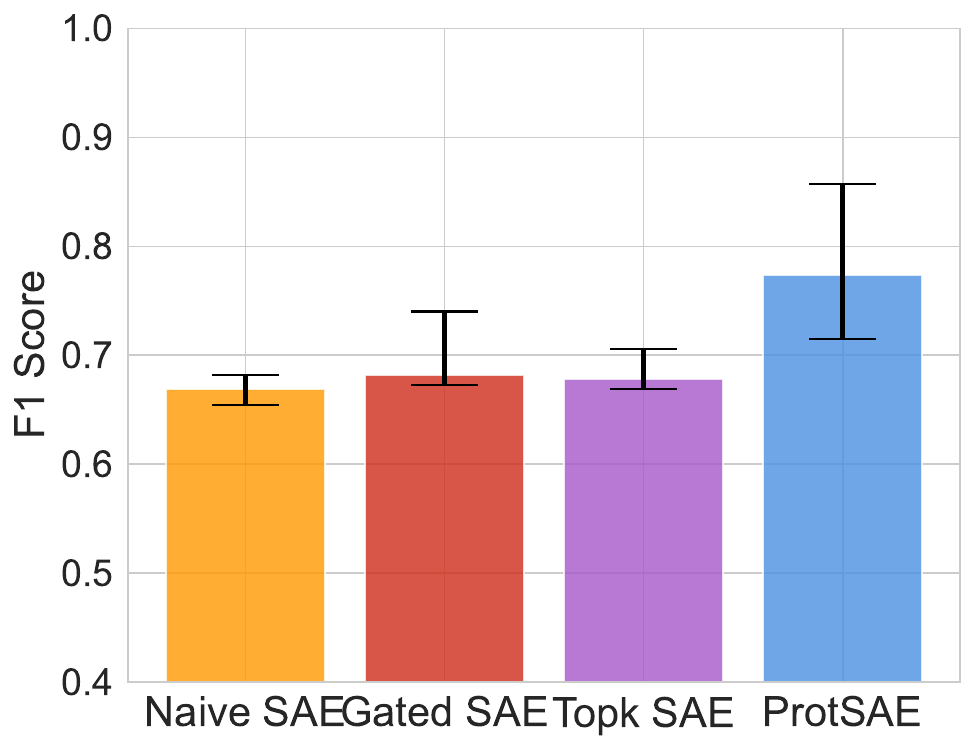}
        \caption{Golgi membrane}
    \end{subfigure}
        \hfill
    \begin{subfigure}[t]{0.21\textwidth}
        \centering
        \includegraphics[width=\linewidth]{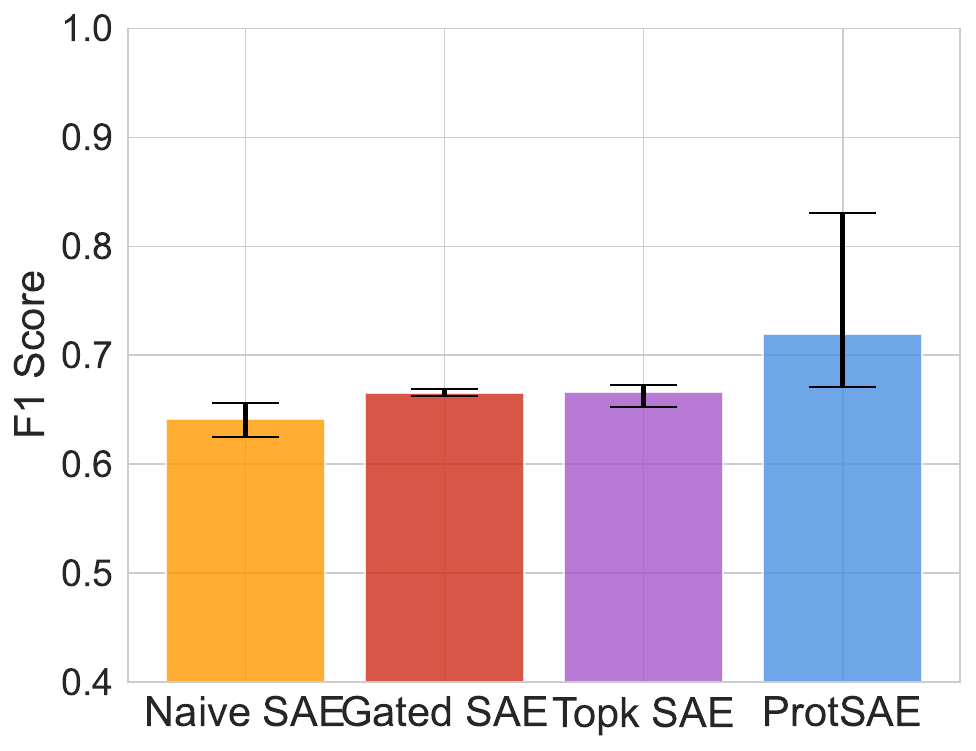}
        \caption{Microtubule binding}
    \end{subfigure}
    \hfill
    \begin{subfigure}[t]{0.21\textwidth}
        \centering
        \includegraphics[width=\linewidth]{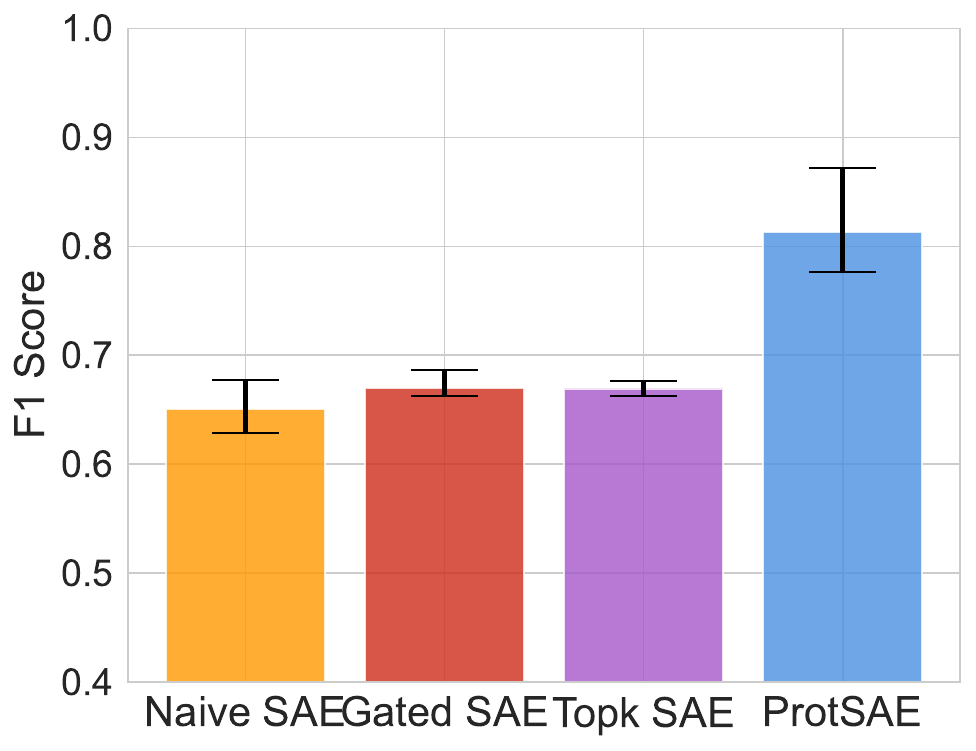}
        \caption{Microtubule bundle formation}
    \end{subfigure}
        \hfill
    \begin{subfigure}[t]{0.21\textwidth}
        \centering
        \includegraphics[width=\linewidth]{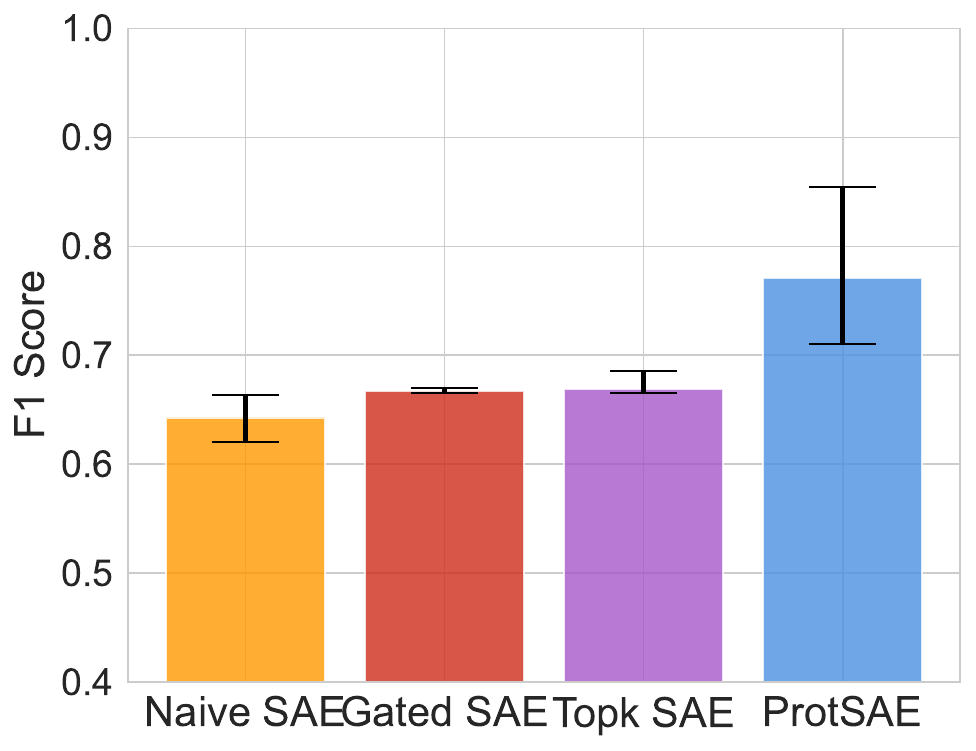}
        \caption{Nuclear chromosome}
    \end{subfigure}
        \hfill
    \begin{subfigure}[t]{0.21\textwidth}
        \centering
        \includegraphics[width=\linewidth]{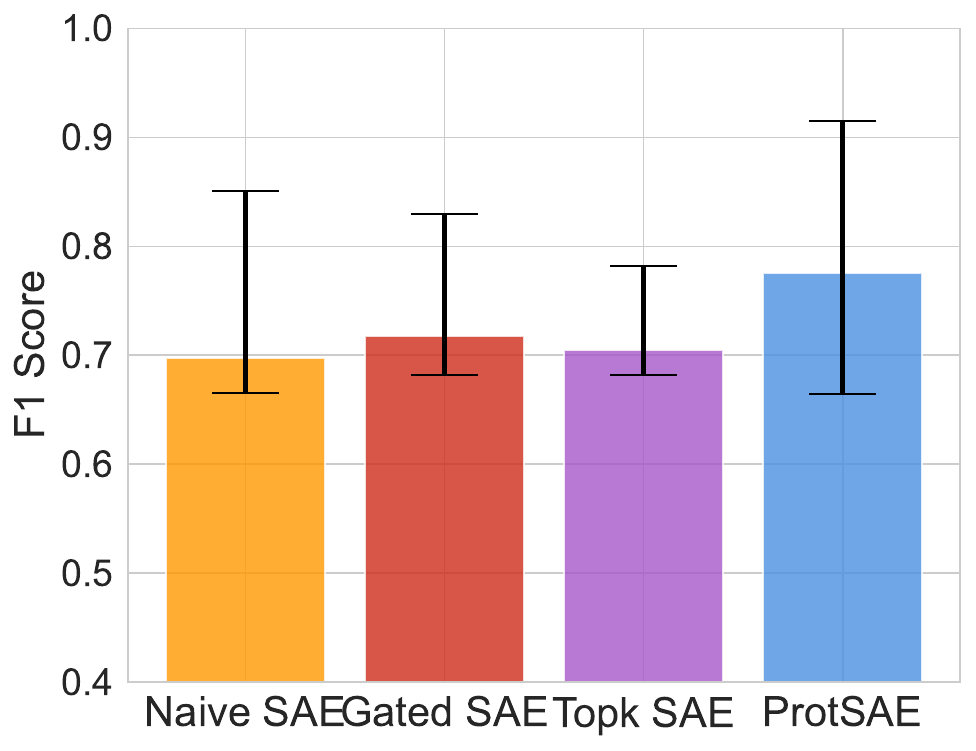}
        \caption{Plastid envelope}
    \end{subfigure}
    \hfill
    \begin{subfigure}[t]{0.21\textwidth}
        \centering
        \includegraphics[width=\linewidth]{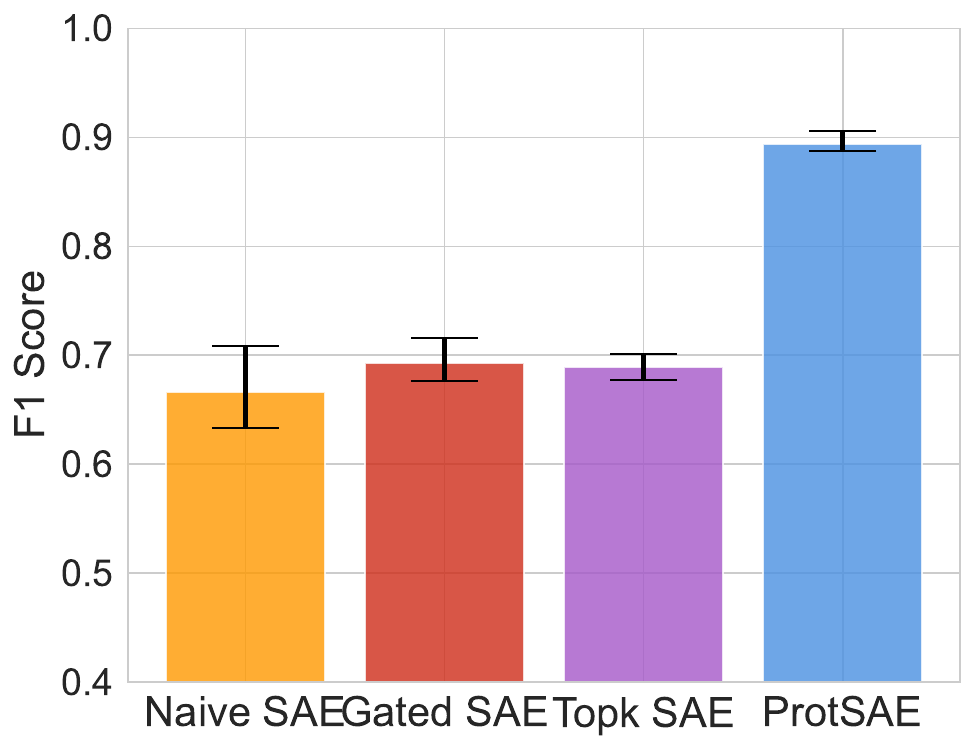}
        \caption{Protein localization to cilium}
    \end{subfigure}
        \hfill
    \begin{subfigure}[t]{0.21\textwidth}
        \centering
        \includegraphics[width=\linewidth]{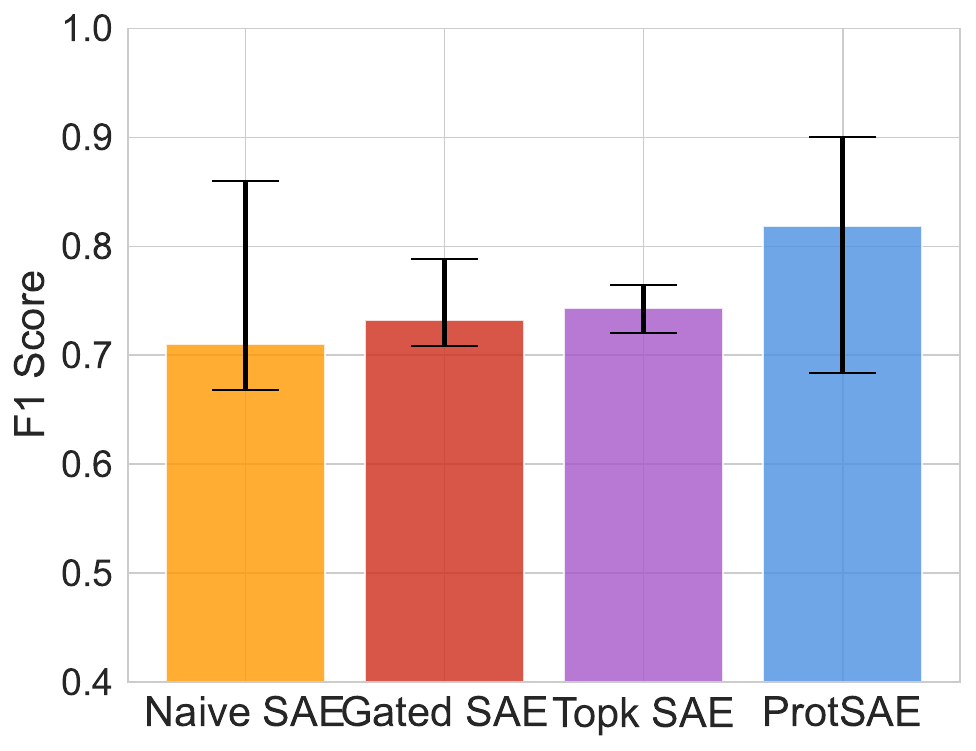}
        \caption{Receptor ligand activity}
    \end{subfigure}
        \hfill
    \begin{subfigure}[t]{0.21\textwidth}
        \centering
        \includegraphics[width=\linewidth]{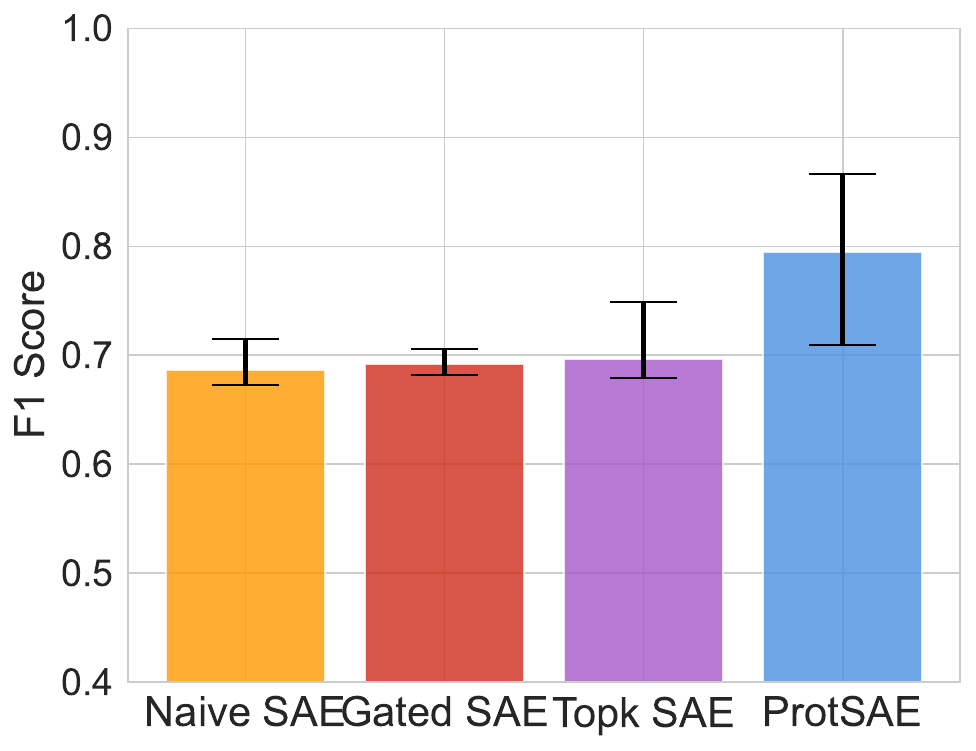}
        \caption{rRNA metabolic process}
    \end{subfigure}
    \hfill
    \begin{subfigure}[t]{0.21\textwidth}
        \centering
        \includegraphics[width=\linewidth]{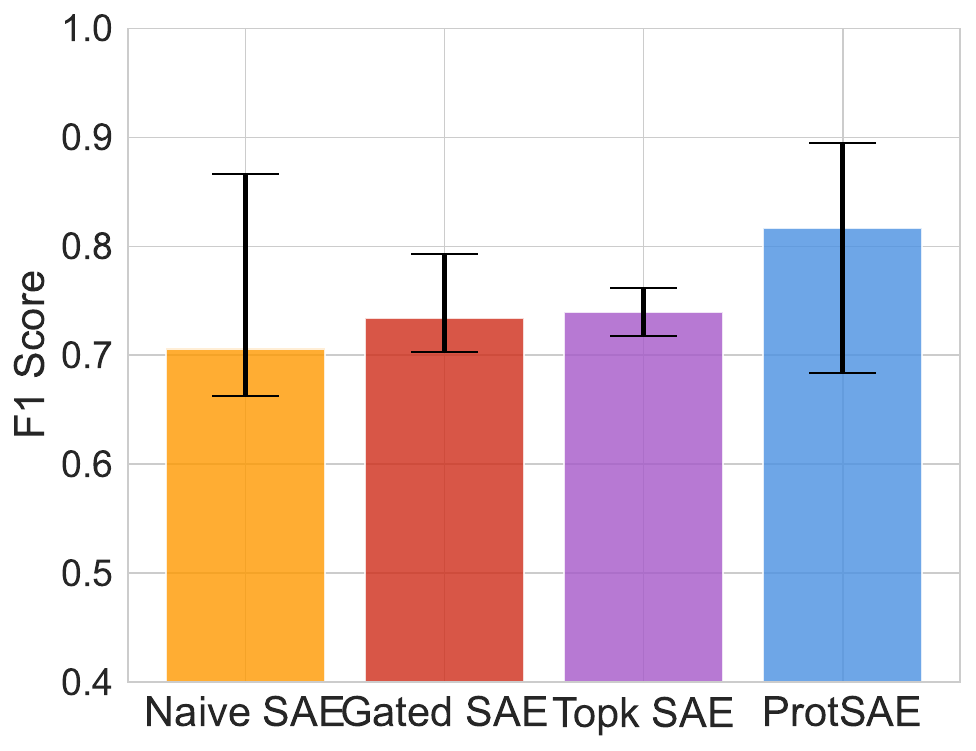}
        \caption{Signaling receptor activator activity}
    \end{subfigure}
        \hfill
    \begin{subfigure}[t]{0.21\textwidth}
        \centering
        \includegraphics[width=\linewidth]{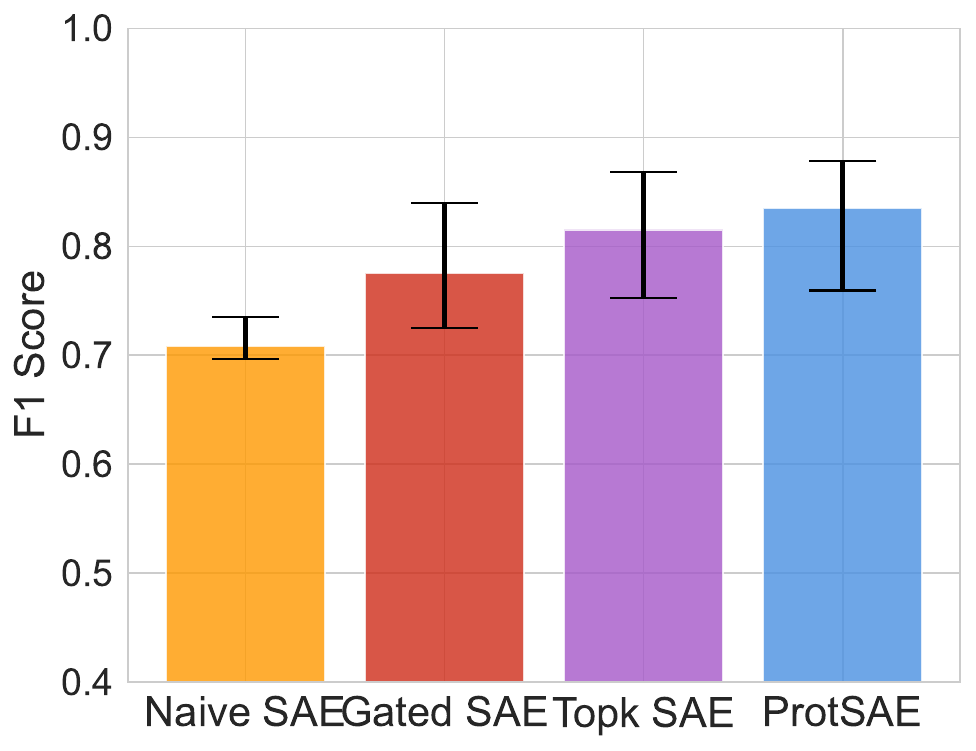}
        \caption{Sodium ion transport}
    \end{subfigure}
        \hfill
    \begin{subfigure}[t]{0.21\textwidth}
        \centering
        \includegraphics[width=\linewidth]{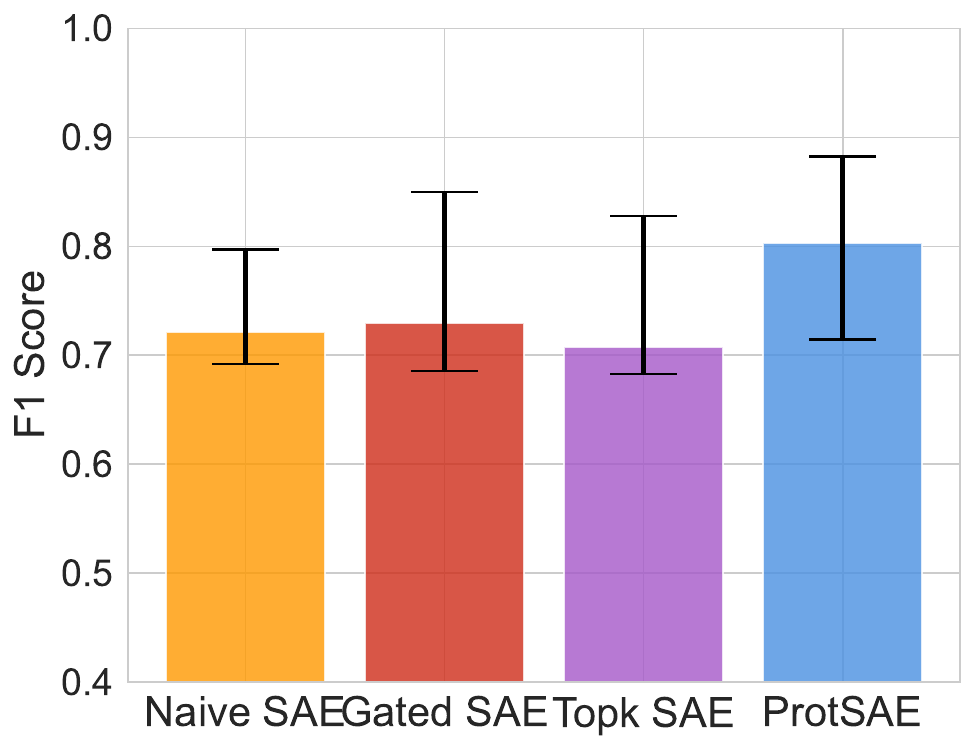}
        \caption{Storage vacuole}
    \end{subfigure}
    \hfill
    \begin{subfigure}[t]{0.21\textwidth}
        \centering
        \includegraphics[width=\linewidth]{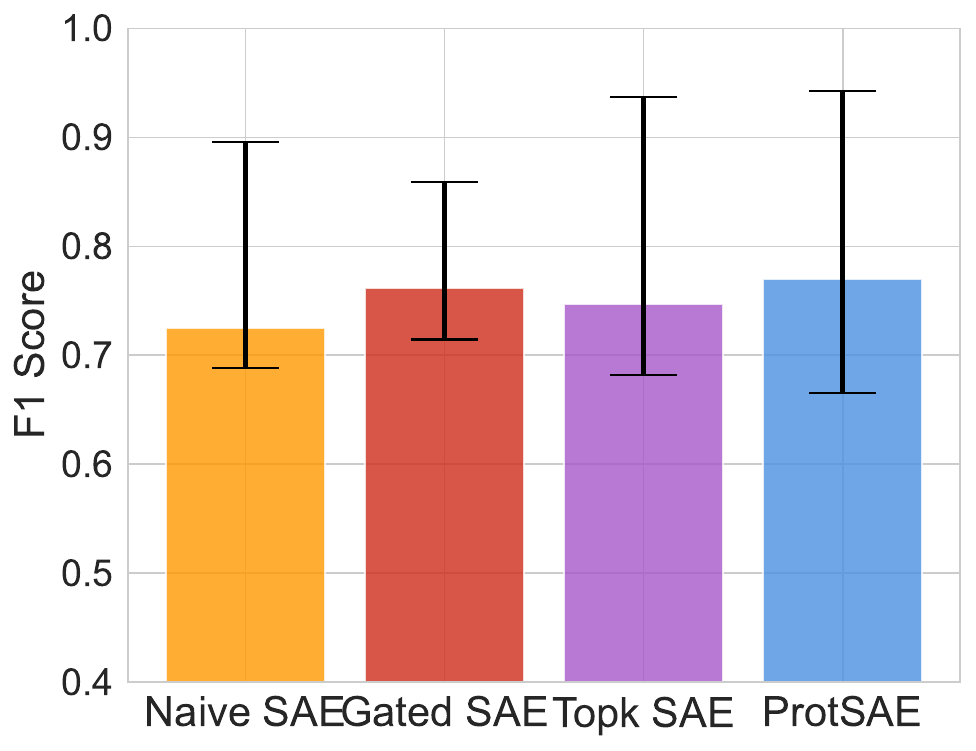}
        \caption{Structural constituent of ribosome}
    \end{subfigure}
        \hfill
    \begin{subfigure}[t]{0.21\textwidth}
        \centering
        \includegraphics[width=\linewidth]{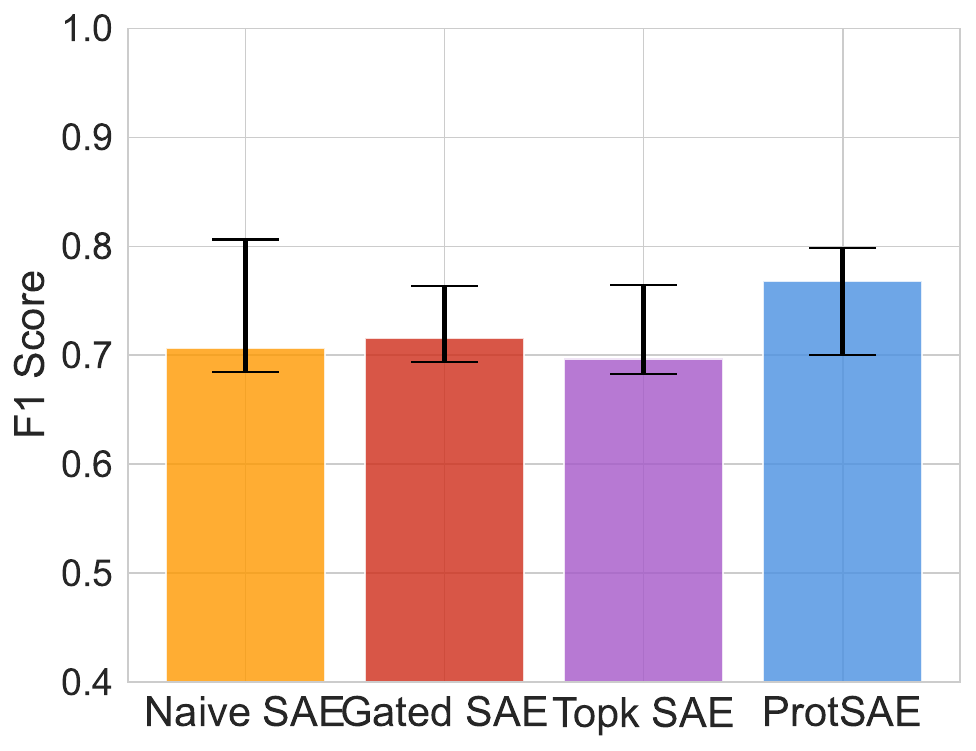}
        \caption{tRNA metabolic process}
    \end{subfigure}
        \hfill
    \begin{subfigure}[t]{0.21\textwidth}
        \centering
        \includegraphics[width=\linewidth]{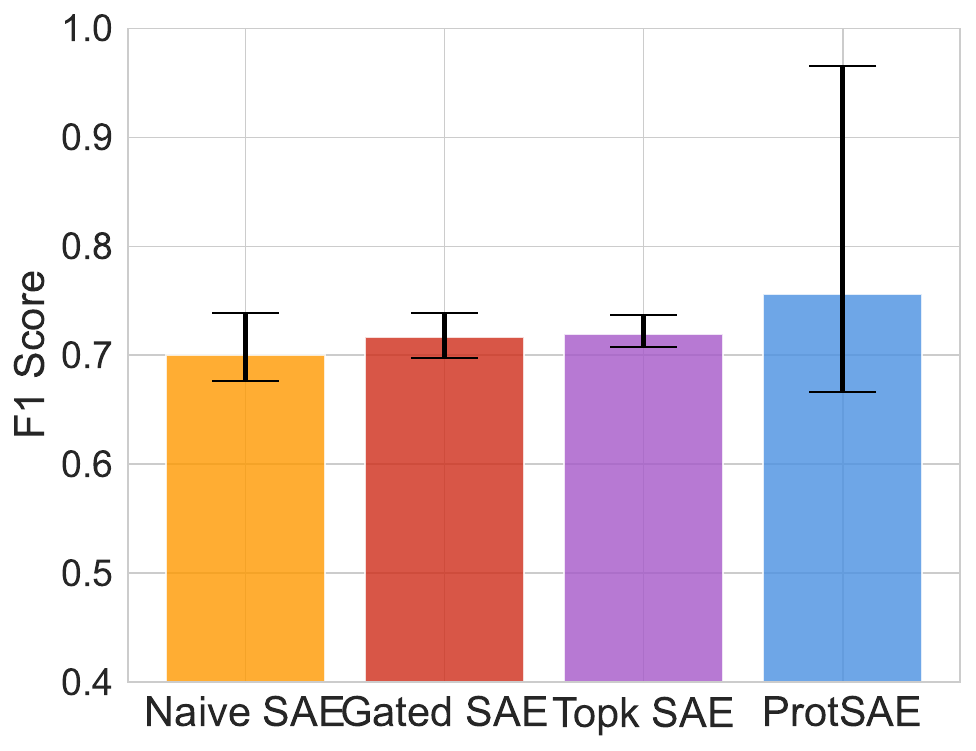}
        \caption{UDP-glycosyltransferase activity}
    \end{subfigure}
    \hfill
    \begin{subfigure}[t]{0.21\textwidth}
        \centering
        \includegraphics[width=\linewidth]{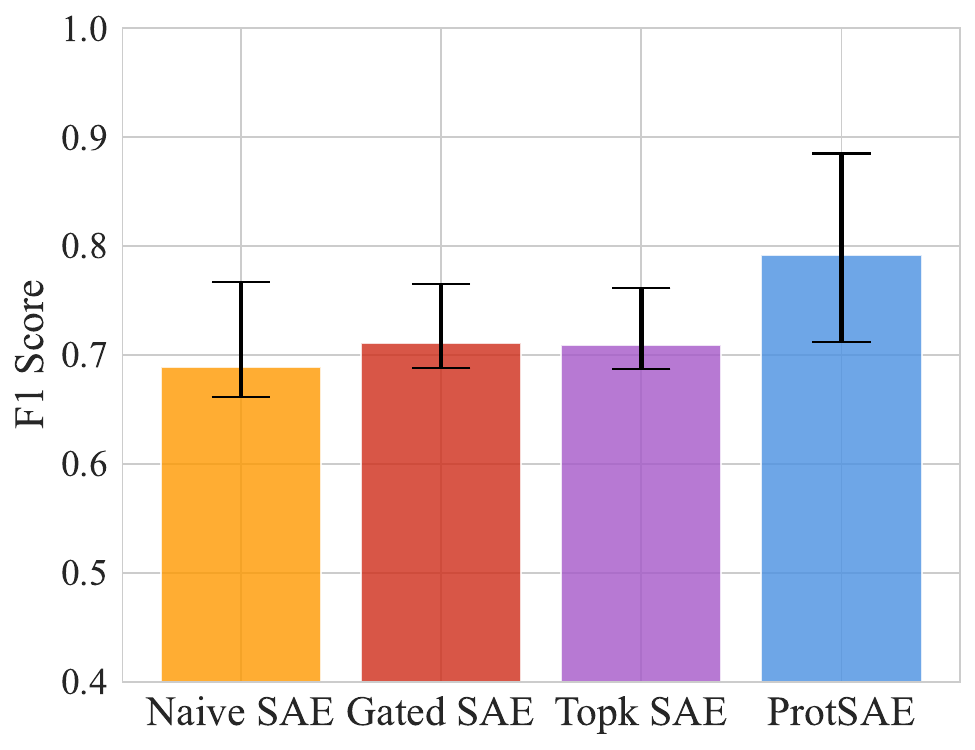}
        \caption{Average performance on 15 concepts}
    \end{subfigure}
    
    \caption{Top-10 activated concept analysis results for 15 GO terms}
    \label{appendix_fig:interpretability_all}
\end{figure*}

\section{Steering Experiment}
\label{appendix:steering_experiments}

\subsection{Settings}
\label{appendix:settings2}

We design intervention experiments based on seven concepts to evaluate whether \modelname captures semantics related to specific biological concepts and consequently influences the generation of PLMs.
The specific GO terms being intervened include: GO:0000793 ``\textit{condensed chromosome}'', GO:0009570 ``\textit{chloroplast stroma}'', GO:0036464 ``\textit{cytoplasmic ribonucleoprotein granule}'', GO:0009617 ``\textit{response to bacterium}'', GO:0016579 ``\textit{protein deubiquitination}'', GO:0032543 ``\textit{mitochondrial translation}'', and GO:0070646 ``\textit{protein modification by small protein removal}''.
For each concept, we construct an evaluation dataset consisting of 5{,}000 samples, and generate 500 samples with intervention.
Following previous works~\cite{taxdiff,prollama,CtrlProt}, we assess the structural similarity between proteins generated after intervention and those in the evaluation dataset. A higher degree of structural similarity indicates that the protein language model, when guided by \modelname, tends to generate proteins structurally aligned with the targeted concept, potentially sharing similar functions. 
We use ESMFold~\cite{esmfold} for structure prediction of the generated protein sequences.

\paragraph{Metrics.} 
We use Foldseek~\cite{foldseek} to evaluate structural similarity using the Template Modeling score (TM-score)~\cite{tm-score} and Root Mean Square Distance (RMSD)~\cite{rmsd}.
A higher TM-score and a lower RMSD, indicates greater structural similarity between the generated proteins and those in the evaluation dataset.
The predicted Local Distance Difference Test (pLDDT) score is used to evaluate the confidence of structure prediction. 
A higher pLDDT reflects more reliable predictions and greater structural stability.

\begin{figure*}[t]
    \centering
    \begin{subfigure}[t]{0.3\textwidth}
        \centering
        \includegraphics[width=\textwidth]{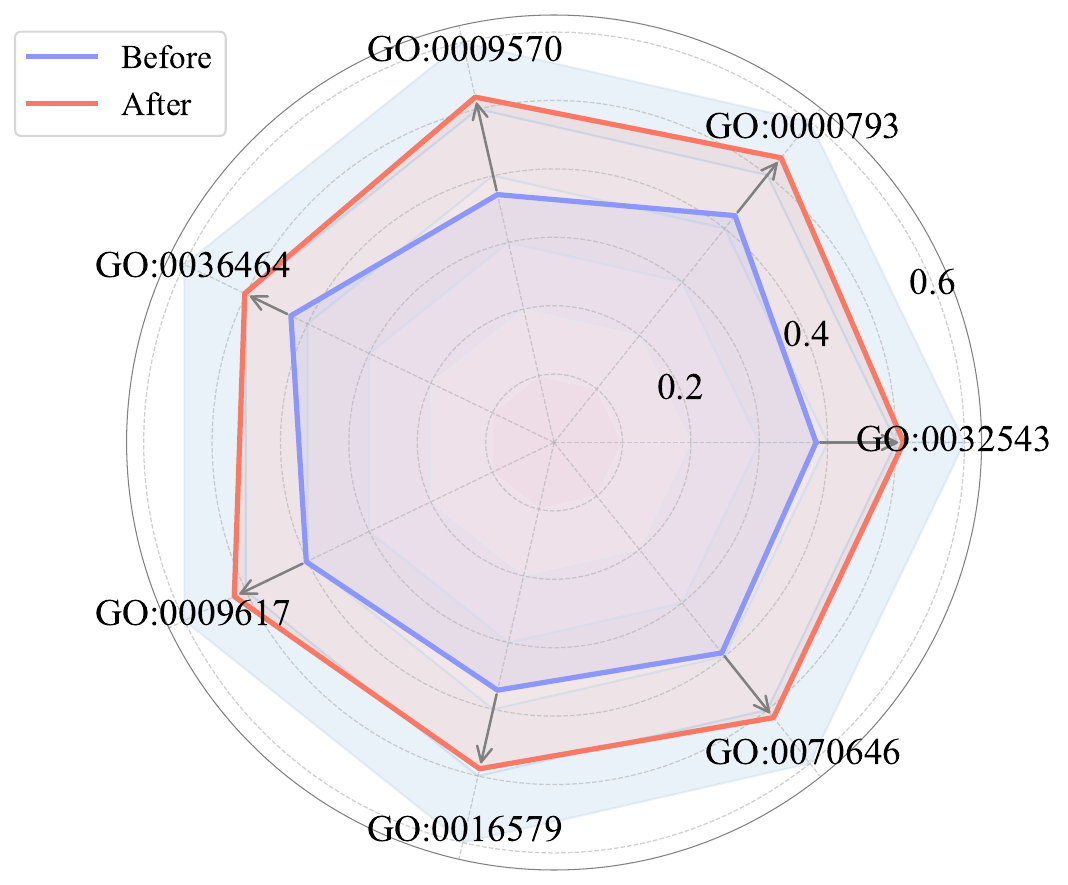}
        \caption{Changes in pLDDT before and after steering}
        \label{appendix_fig:steering_plddt}
    \end{subfigure}
    \begin{subfigure}[t]{0.3\textwidth}
        \centering
        \includegraphics[width=\textwidth]{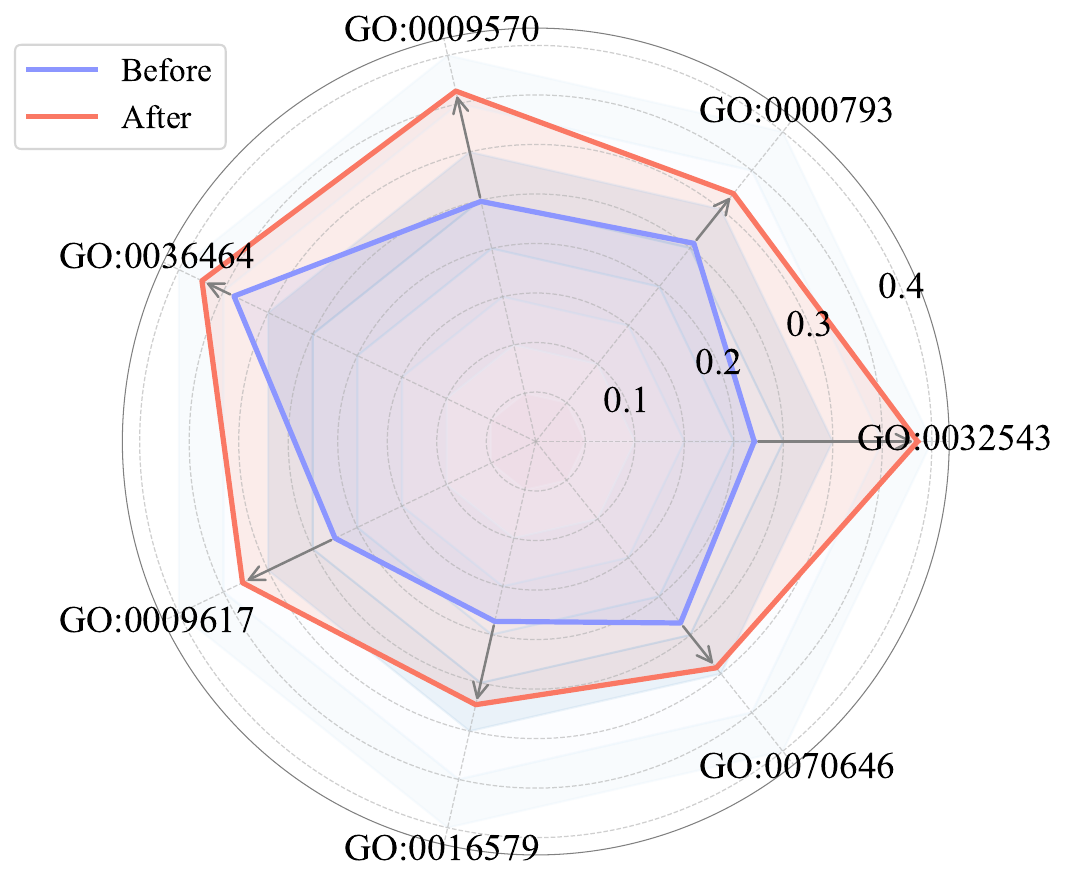}
        \caption{Changes in TM-score before and after steering}
        \label{appendix_fig:steering_tmscore}
    \end{subfigure}
    \begin{subfigure}[t]{0.3\textwidth}
        \centering
        \includegraphics[width=\textwidth]{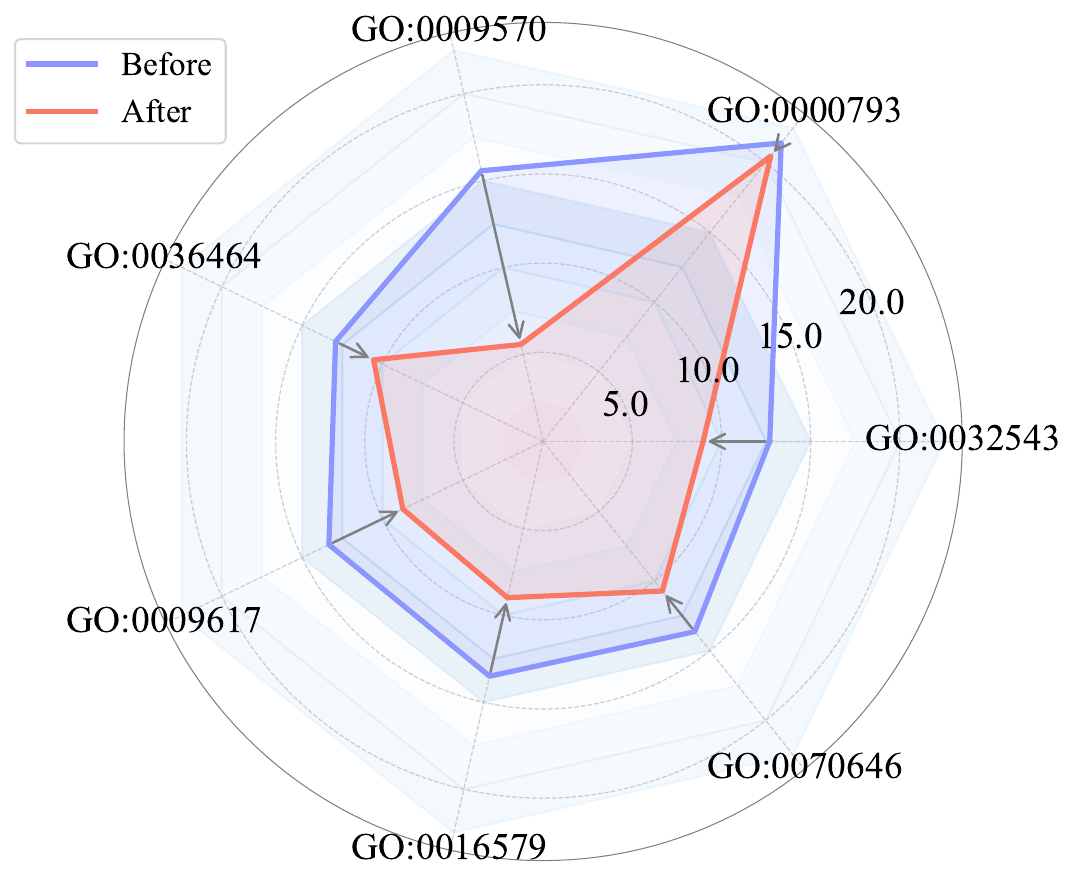}
        \caption{Changes in RMSD before and after steering}
        \label{appendix_fig:steering_rmsd}
    \end{subfigure}
    
    \caption{Detailed steering performance}
    \label{appendix_fig:steering_results}
\end{figure*}

\begin{table*}
\centering
{\small
\begin{tabular}{lccccccccc}
\toprule
Method & Dataset &  $L_0$ &  Proportion &   MSE &   AUC &  $F_{\max}$ &  $S_{\min}$ &  AUPR &  Recovered \\
\midrule
Naive SAE &      MFO &  329.6 &        1.00 & 0.365 & 0.473 &       0.332 &      14.686 & 0.239 &      0.804 \\
Naive SAE &      MFO &  158.3 &        1.00 & 0.502 & 0.478 &       0.334 &      14.933 & 0.245 &      0.698 \\
Naive SAE &      MFO &   75.7 &        1.00 & 0.642 & 0.478 &       0.335 &      15.159 & 0.246 &      0.560 \\
Naive SAE &      MFO & 1081.7 &        1.00 & 0.151 & 0.468 &       0.327 &      14.578 & 0.225 &      0.905 \\
Naive SAE &      MFO &  895.6 &        1.00 & 0.183 & 0.467 &       0.329 &      14.574 & 0.241 &      0.902 \\
\midrule
Gated SAE &      MFO & 1052.8 &        1.00 & 0.121 & 0.483 &       0.351 &      14.292 & 0.259 &      0.953 \\
Gated SAE &      MFO &  840.6 &        1.00 & 0.151 & 0.482 &       0.400 &      14.314 & 0.254 &      0.938 \\
Gated SAE &      MFO &  561.3 &        1.00 & 0.221 & 0.481 &       0.667 &      14.361 & 0.261 &      0.889 \\
Gated SAE &      MFO &  381.6 &        1.00 & 0.315 & 0.478 &       0.340 &      14.439 & 0.233 &      0.853 \\
Gated SAE &      MFO &  265.0 &        1.00 & 0.384 & 0.481 &       0.335 &      14.522 & 0.230 &      0.792 \\
\midrule
TopK SAE &      MFO &   50.0 &        1.00 & 0.471 & 0.508 &       0.500 &      14.177 & 0.264 &      0.687 \\
TopK SAE &      MFO &  100.0 &        1.00 & 0.385 & 0.522 &       0.387 &      14.166 & 0.271 &      0.763 \\ 
TopK SAE &      MFO &  500.0 &        1.00 & 0.196 & 0.523 &       0.364 &      14.140 & 0.268 &      0.928 \\
TopK SAE &      MFO & 1000.0 &        1.00 & 0.111 & 0.529 &       0.355 &      14.131 & 0.277 &      0.968 \\
\midrule
\modelname &      MFO &   50.0 &        0.85 & 0.493 & 0.700 &       0.591 &      12.645 & 0.441 &      0.692 \\
\modelname &      MFO &  100.0 &        1.00 & 0.404 & 0.747 &       0.591 &      12.352 & 0.434 &      0.770 \\
\modelname &      MFO &  500.0 &        1.00 & 0.203 & 0.789 &       0.675 &      12.209 & 0.441 &      0.914 \\
\modelname &      MFO & 1000.0 &        1.00 & 0.115 & 0.802 &       0.646 &      12.205 & 0.439 &      0.955 \\
\bottomrule
\end{tabular}}
\caption{Detailed results on the Molecular Function Ontology dataset}
\label{appendix_tab:mf_metrics}
\end{table*}

\paragraph{Intervention method.} 
\label{appendix:intervention_method}
Let $p = \{t_1, t_2, \dots, t_L\}$ be a protein sequence of length $L$, where $t_j$ denotes the token at position $j$.
Given a target concept $c_i$, we compute the relevance score $\pi_{\text{pred}}^{(i, j)}$ for each token $t_j$ with respect to $c_i$, from Eq.~\eqref{eq:prediction}.
We define a masking threshold $\theta_i$, as the 50\%-position value in the sorted predicted importance scores. 
Then, we construct the masked sequence $\tilde{p}_{c_i}$ by replacing tokens whose importance scores are below the threshold:
\begin{align}
\theta_i &= \operatorname{Median} \left( \left\{ \pi_{\text{pred}}^{(i, j)}\right\}_{j=1}^L\right),\\
\tilde{p}_{c_i} &= \left\{ t_j' \mid t_j' = 
\begin{cases}
\texttt{[MASK]}, & \text{if } \pi_{\text{pred}}^{(i, j)} < \theta_i \\
t_j, & \text{otherwise}
\end{cases}
\right\}.
\end{align}

We use $\tilde{p}_{c_i}$ into the protein language model (e.g., ESM2-15B) to obtain a reconstructed sequence $\hat{p}$.
We simultaneously intervene on the target concept and its ancestor concepts.
We observe that overly strong interventions may disrupt the semantic representations of PLMs.
To mitigate this, for each concept, we enhance the activation by adding 1.2 times the mean of its Top-$K$ activations.

\subsection{Detailed Steering Results}
\label{appendix:detailed_steering_results}
Here we show the average results before and after intervention on TM-score, RMSD and pLDDT in Figure~\ref{appendix_fig:steering_results}.
We observe that, after intervention, the generated samples exhibit significantly improved structural similarity to the evaluation set.
At the same time, the structural stability of the generated proteins also increases accordingly.
This indicates that \modelname successfully captures protein structural features associated with the target function and can effectively guide the generation process through semantic intervention.

\section{Limitations}
\label{appendix:limitation}
Currently, we evaluate \modelname on the ESM2 protein language model, and it has not been scaled to broader architectures or application scenarios. 
In future work, we aim to extend \modelname for more extensive exploration.
For example, \modelname can be trained on the protein language models aligned with natural language.
It enables us to investigate the association between natural language descriptions and structural fragments within protein sequences.
\modelname can also be trained on the sequence-structure co-design model.
Moreover, we can further explore how to extract interpretable features from protein structure prediction models and generative diffusion models.


\section{Tables of All Training Results}
\label{appendix:tablesofresults}
Here, we report the detailed training results of \modelname compared to SAE baselines across three datasets in Tables~\ref{appendix_tab:mf_metrics}, \ref{appendix_tab:bp_metrics}, and \ref{appendix_tab:cc_metrics}.

\begin{table*}
\centering
{\small
\begin{tabular}{lcrccccccc}
\toprule
Method & Dataset &  $L_0$ &  Proportion &   MSE &   AUC &  $F_{\max}$ &  $S_{\min}$ &  AUPR &  Recovered \\
\midrule
Naive SAE &      BPO &  443.8 &        1.00 & 0.302 & 0.516 &       0.306 &      43.743 & 0.236 &      0.872 \\
Naive SAE &      BPO &  230.6 &        1.00 & 0.423 & 0.510 &       0.305 &      44.378 & 0.237 &      0.785 \\
Naive SAE &      BPO &  108.1 &        1.00 & 0.571 & 0.506 &       0.305 &      44.223 & 0.237 &      0.659 \\
Naive SAE &      BPO & 1281.5 &        1.00 & 0.122 & 0.532 &       0.305 &      43.566 & 0.228 &      0.987 \\
Naive SAE &      BPO & 1080.4 &        1.00 & 0.147 & 0.527 &       0.305 &      43.570 & 0.224 &      0.971 \\
\midrule
Gated SAE &      BPO &  984.5 &        1.00 & 0.138 & 0.546 &       0.328 &      42.712 & 0.248 &      0.972 \\
Gated SAE &      BPO &  700.4 &        1.00 & 0.175 & 0.542 &       0.326 &      42.775 & 0.249 &      0.933 \\
Gated SAE &      BPO &  520.0 &        1.00 & 0.216 & 0.538 &       0.322 &      42.882 & 0.247 &      0.909 \\
Gated SAE &      BPO &  375.8 &        1.00 & 0.257 & 0.533 &       0.318 &      43.040 & 0.243 &      0.875 \\
Gated SAE &      BPO &  260.0 &        1.00 & 0.337 & 0.524 &       0.313 &      43.250 & 0.241 &      0.807 \\
\midrule
TopK SAE &      BPO &   50.0 &        1.00 & 0.461 & 0.517 &       0.324 &      42.850 & 0.257 &      0.722 \\
TopK SAE &      BPO &  100.0 &        1.00 & 0.374 & 0.532 &       0.326 &      42.773 & 0.252 &      0.800 \\
TopK SAE &      BPO &  500.0 &        1.00 & 0.186 & 0.536 &       0.324 &      42.792 & 0.249 &      0.929 \\
TopK SAE &      BPO & 1000.0 &        1.00 & 0.104 & 0.546 &       0.325 &      42.800 & 0.246 &      0.992 \\
\midrule
\modelname &      BPO &   50.0 &        0.93 & 0.501 & 0.662 &       0.380 &      41.365 & 0.315 &      0.683 \\
\modelname &      BPO &  100.0 &        0.99 & 0.397 & 0.684 &       0.381 &      41.249 & 0.316 &      0.751 \\
\modelname &      BPO &  500.0 &        1.00 & 0.200 & 0.736 &       0.385 &      40.992 & 0.322 &      0.924 \\
\modelname &      BPO & 1000.0 &        1.00 & 0.109 & 0.753 &       0.393 &      40.728 & 0.331 &      0.980 \\
\bottomrule
\end{tabular}}
\caption{Detailed results on the Biological Process Ontology dataset}
\label{appendix_tab:bp_metrics}
\end{table*}

\begin{table*}
\centering
{\small
\begin{tabular}{lccccccccc}
\toprule
Method & Dataset &  $L_0$ &  Proportion &   MSE &   AUC &  $F_{\max}$ &  $S_{\min}$ &  AUPR &  Recovered \\
\midrule
Naive SAE &      CCO &  434.6 &         1.0 & 0.303 & 0.528 &       0.631 &      11.824 & 0.572 &      0.837 \\
Naive SAE &      CCO &  224.3 &         1.0 & 0.425 & 0.522 &       0.630 &      11.955 & 0.570 &      0.758 \\
Naive SAE &      CCO &  111.7 &         1.0 & 0.560 & 0.524 &       0.625 &      12.000 & 0.600 &      0.632 \\
Naive SAE &      CCO & 1228.8 &         1.0 & 0.126 & 0.541 &       0.631 &      11.744 & 0.571 &      0.970 \\
Naive SAE &      CCO & 1040.9 &         1.0 & 0.153 & 0.539 &       0.630 &      11.793 & 0.571 &      0.943 \\
\midrule
Gated SAE &      CCO &  938.1 &         1.0 & 0.134 & 0.570 &       0.643 &      11.224 & 0.598 &      0.965 \\
Gated SAE &      CCO &  663.6 &         1.0 & 0.189 & 0.559 &       0.641 &      11.292 & 0.596 &      0.934 \\
Gated SAE &      CCO &  492.0 &         1.0 & 0.229 & 0.536 &       0.641 &      11.387 & 0.590 &      0.908 \\
Gated SAE &      CCO &  358.7 &         1.0 & 0.296 & 0.534 &       0.638 &      11.470 & 0.584 &      0.852 \\
Gated SAE &      CCO &  257.9 &         1.0 & 0.353 & 0.527 &       0.636 &      11.611 & 0.581 &      0.803 \\
\midrule
TopK SAE &      CCO &   50.0 &         1.0 & 0.460 & 0.595 &       0.651 &      11.170 & 0.608 &      0.718 \\
TopK SAE &      CCO &  100.0 &         1.0 & 0.373 & 0.625 &       0.653 &      11.067 & 0.615 &      0.783 \\
TopK SAE &      CCO &  500.0 &         1.0 & 0.186 & 0.617 &       0.651 &      11.131 & 0.611 &      0.916 \\
TopK SAE &      CCO & 1000.0 &         1.0 & 0.104 & 0.621 &       0.651 &      11.110 & 0.614 &      0.991 \\
\midrule
\modelname &      CCO &   50.0 &         1.0 & 0.497 & 0.771 &       0.694 &      10.044 & 0.688 &      0.682 \\
\modelname &      CCO &  100.0 &         1.0 & 0.397 & 0.793 &       0.696 &       9.978 & 0.692 &      0.773 \\
\modelname &      CCO &  500.0 &         1.0 & 0.200 & 0.829 &       0.697 &       9.951 & 0.690 &      0.916 \\
\modelname &      CCO & 1000.0 &         1.0 & 0.108 & 0.835 &       0.698 &       9.885 & 0.690 &      0.987 \\ 
\bottomrule
\end{tabular}}
\caption{Detailed results on the Cellular Component Ontology dataset}
\label{appendix_tab:cc_metrics}
\end{table*}

\end{document}